\documentclass[reprint, superscriptaddress, aps, floatfix, prb]{revtex4-2}

\usepackage[nice]{nicefrac}
\usepackage{graphicx,color}
\usepackage{amsmath,amssymb,bm,bbm}
\usepackage{amsmath}
\usepackage{braket}
\usepackage{bbold}
\usepackage{float}
\usepackage{placeins}
\usepackage{soul}
\usepackage{enumitem}
\usepackage{mathtools}
\usepackage{todonotes}
\usepackage{subdepth}

\usepackage{comment}
\usepackage[T1]{fontenc}

\usepackage[plainpages=false,pdfpagelabels,colorlinks=true,linkcolor=red,urlcolor=blue,citecolor=blue,pdftitle={},pdfauthor={},pdfdisplaydoctitle=true,pdfduplex=DuplexFlipLongEdge]{hyperref}

\definecolor{darkred}{rgb}{0.90,0.2,0.2}
\definecolor{darkgreen}{rgb}{0,0.60,.2}
\definecolor{darkblue}{rgb}{0.1,0.3,1}
\definecolor{grey}{cmyk}{0,0,0,0.25}
\definecolor{orange}{cmyk}{0,0.6,0.8,0}

\begin{document}

\title{Two-Parameter Ansatz for the Violation of Eigenstate Thermalization}

\author{Kohei Ohgane}
\affiliation{Department of Theoretical Physics, J. Stefan Institute, SI-1000 Ljubljana, Slovenia}
\author{Lev Vidmar}
\affiliation{Department of Theoretical Physics, J. Stefan Institute, SI-1000 Ljubljana, Slovenia}
\affiliation{Department of Physics, Faculty of Mathematics and Physics, University of Ljubljana, SI-1000 Ljubljana, Slovenia\looseness=-1}

\begin{abstract}
The eigenstate thermalization hypothesis (ETH) provides the prevailing framework for understanding quantum thermalization and ergodicity in isolated many-body systems. 
Yet, no general theory describes the continuous onset of ETH violation between the conventional ETH and its complete breakdown.
Here, we introduce a two-parameter ansatz for ETH violation that unifies and distinguishes two mechanisms: fading ergodicity and trapped ergodicity. While fading ergodicity captures the established route to ergodicity breaking, trapped ergodicity describes a distinct scenario in which ETH is violated in finite systems but restored in the thermodynamic limit. We test this framework in the spin-1/2 $J_1$-$J_2$ chains with onsite disorder and a linear potential. In both cases, we find that the observed ETH violation is consistent with trapped ergodicity. 
\end{abstract}
\maketitle

\section{Introduction}

The ETH is a successful theory for describing quantum thermalization and ergodicity in isolated quantum systems~\cite{deutsch_91, srednicki_94, dalessio_kafri_16}. Over the past few decades, the ETH has been numerically demonstrated for a wide range of quantum lattice models~\cite{rigol_dunjko_08, rigol09, rigol_09a, santos_rigol_10b, rigol_santos_10, steinigeweg_herbrych_13, khatami_pupillo_13, torresherrera_santos_14, beugeling_moessner_14, sorg14, steinigeweg_khodja_14, kim_ikeda_14, khodja_steinigeweg_15, beugeling_moessner_15, dalessio_kafri_16, mondaini_fratus_16, lan_powell_17, lan_powell_17, mondaini_rigol_17, nation_porras_18, yoshizawa_iyoda_18, jansen_stolpp_19, khaymovich_haque_19, leblond_mallayya_19, Mierzejewski_2020, brenes_leblond_20, brenes_goold_20, leblond_rigol_20, richter_dymarsky_20, wang_lamann_22, noh_21, schoenle_jansen_21, sugimoto_hamazaki_21, noh_23, wang_zhu_24, luo_trivedi_24, ebner_schafer_24, patil_rigol_25, capizzi_wang_25, saiaramthottil_emamikopaei_25}.
Its key feature is the strong suppression of eigenstate-to-eigenstate fluctuations of the observable matrix elements above the smooth function, which decay as the inverse square-root of the many-body density of states $\rho$, i.e., exponentially in system size $L$.
Such behavior is described within the Srednicki ansatz~\cite{srednicki_99}, and we refer to it as {\it conventional} ETH.
Recent research also explored to what extent the observable matrix elements may be considered as uncorrelated~\cite{murthy_srednicki_19b, richter_dymarsky_20, brenes_pappalardi_21, wang_lamann_22, pappalardi_foini_22, hahn_luitz_24, fava_kurchan_25, pappalardi_fritzsch_25}, which is important for the description of quantities beyond two-point correlations~\cite{foini_kurchan_19, chan_deluca_19}.

Soon after the relevance of ETH for paradigmatic quantum lattice models was recognized~\cite{rigol_dunjko_08}, significant effort was devoted to identifying counterexamples to the Srednicki ansatz. Two main classes of ETH breakdown emerged.
In the first class, the variance of the diagonal matrix-element fluctuations (above the smooth function) decays polynomially with system size $L$, as in quadratic systems~\cite{biroli_kollath_10, Cassidy_2011, vidmar16, zhang_vidmar_22, lydzba_swietek_24} and integrable interacting systems~\cite{steinigeweg_herbrych_13, Ikeda2013, beugeling_moessner_14, alba_15, mori_16, leblond_mallayya_19, Mierzejewski_2020, essler_deklerk_24, lydzba_swietek_24}, or even remains nonzero in the thermodynamic limit, as occurs for certain observables in localized quadratic systems~\cite{lydzba_swietek_24}.
We refer to the later as the {\it complete breakdown} of the ETH.
In the second class, a measure-zero subset of outlier matrix elements deviates from the corresponding microcanonical averages, while the variance of the matrix-element fluctuations still decays exponentially with $L$, as observed in systems exhibiting quantum many-body scars~\cite{shiraishi_mori_17, turner_michailidis_18b, dooley_kells_22, wang_zhou_25} or Hilbert-space fragmentation~\cite{moudgalya_prem_21, aditya_dhar_24}.
However, neither scenario describes the ETH violation as a continuous process interpolating between the conventional ETH regime governed by the Srednicki ansatz and a complete breakdown of ETH.

\begin{figure}[!t]
\includegraphics[width=1.0\columnwidth]{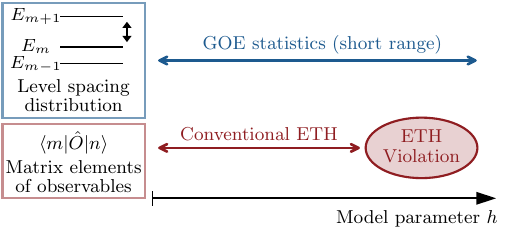}
\vspace{-0.2cm}
\caption{ 
The ETH breakdown in finite systems upon tuning a model parameter $h$ away from the ergodic regime. Deep in the ergodic regime, observable matrix elements are described by the conventional ETH. We focus on the boundary of validity of the conventional ETH, where short-range spectral statistics, such as the level-spacing distribution, remain consistent with GOE predictions.
}
\label{fig:sketch}
\end{figure}

In this work, we fill this gap by introducing a two-parameter ansatz for ETH violation that smoothly interpolates between the conventional ETH and the complete breakdown of the ETH. 
We focus in particular on the regime at the boundary of validity of the conventional ETH, where numerical results no longer support a straightforward application of the Srednicki ansatz, while short-range spectral statistics remain consistent with GOE predictions (see Fig.~\ref{fig:sketch}). 
As suggested in Ref.~\cite{kliczkowski_swietek_24}, such a sequence of breakdowns---first of the conventional ETH and only subsequently of short-range GOE statistics---may be a generic feature of quantum many-body systems near the boundary of ergodicity.
We conjecture that this intermediate regime contains key information about the nature of the ergodicity breaking that emerges upon further departure from the ergodic regime in finite systems.

The two-parameter ansatz for ETH violation allows us to distinguish two scenarios.
The first scenario is {\it fading ergodicity}, which has recently been introduced as a precursor regime of an ergodicity breaking phase transition~\cite{kliczkowski_swietek_24, swietek_lydzba_25, swietek_hopjan_25, swietek_kliczkowski_26}.
The second scenario, which is the new result of this work, is dubbed {\it trapped ergodicity}.
It corresponds to the finite-size regime in which conventional ETH emerges only when the system size $L$ exceeds a characteristic length scale $\xi$, dubbed the ergodization length.
Hence, ergodicity appears to be trapped by finite-size effects. 
We numerically study spin-1/2 $J_1$-$J_2$ chains with onsite disorder and a constant electric field, and show that trapped ergodicity characterizes the ETH violation in these models.

\section{Conventional ETH}

The conventional ETH was formulated by Srednicki~\cite{srednicki_99} as an ansatz for the matrix elements $A_{mn} \equiv \bra{m}\hat{A}\ket{n}$ of an observable $\hat{A}$ in Hamiltonian eigenstates $\{|n\rangle\}$, with $\hat H|n\rangle = E_n|n\rangle$,
\begin{eqnarray}
    A_{mn} = \mathcal{A}(\bar{E})\delta_{mn} + \rho(\bar{E})^{-1/2} f(\bar{E}, \omega)R_{mn}\,,  \label{eq:ETH}
\end{eqnarray}
where $\bar{E} = (E_{m}+E_{n})/2$ is the mean energy, $\omega = E_{m}-E_{n}$ is the energy difference, and $R_{mn}$ are random numbers with zero mean and unit variance.
$\mathcal{A}(\bar{E})$ and $f(\bar{E}, \omega)$ are smooth functions of the energies, and $\rho(\bar E)$ is the many-body density of states. 
We restrict our discussion to an energy window centered at the mean energy $E_0 = \mathrm{Tr}[\hat{H}]/\mathcal{D}$, where $\cal D$ is the Hilbert-space dimension. 
Accordingly, we omit the explicit dependence on $\bar{E}$ in the functions in Eq.~\eqref{eq:ETH} and denote them as $\rho(\bar{E}) = \rho$ and $f(\bar{E},\omega) = f(\omega)$.
Still, it is known that $\rho$ also depends on the system size $L$, and even $f(\omega)$ may depend on $L$ at $\omega\to 0$; the scaling properties of $\rho$ and $f(\omega)$ with $L$ will be our main focus below.

We study the ETH violation via the fluctuations of the diagonal matrix elements $A_{mm}$, quantified by the corresponding variance~\cite{beugeling_moessner_14, Mierzejewski_2020},
\begin{eqnarray}
    \sigma^2 =  \Bigl\langle  \bigl\{A_{mm} - \mathrm{MA}[A_{mm}]\bigr\}^{2} \Bigr\rangle_{E_0, \delta E}. \label{eq:def_diag_fluc}
\end{eqnarray}
In the numerical calculations, $\langle \cdot \rangle_{E_0, \delta E}$ denotes the microcanonical average over eigenstates $m$ within an energy window of width $\delta E$, centered at $E_0$. 
Here, $\mathrm{MA}[\cdot]$ represents a moving average of the series $\{A_{mm}\}$ centered at $A_{mm}$, defined as $\mathrm{MA}[A_{mm}] = \sum_{n=m-M}^{m+M}A_{nn}/(2M+1)$, where $M$ is an integer.
The role of the moving average is to subtract the impact of the smooth function since  $\mathrm{MA}[A_{mm}] \to \mathcal{A}(E_{m})$.
Hence, $\sigma^2$ can be expressed within the framework of Eq.~\eqref{eq:ETH} as
\begin{equation}
    \sigma^2 = \rho^{-1}\,|f_0|^{2}\,, \label{eq:diag_fluc_ETH}
\end{equation}
where $|f_0|^2 \equiv |f(\omega=0)|^{2}$. 
Equation~\eqref{eq:diag_fluc_ETH} is our starting point for the characterization of ETH violation.

First, we parametrize the dependence of the density of states $\rho$ on the system size $L$ as $\rho = \rho(L) = \bar\rho(L) \exp\{L/\eta_0\}$.
The term $\exp\{L/\eta_0\}$ reflects the exponential increase of the Hilbert-space dimension ${\cal D}$ in interacting quantum systems, and we set $\eta_0 = 1/\ln(2)$, having in mind systems of coupled qubits.
The term $\bar\rho(L)$ is a subleading correction that scales at most polynomially with $L$.
It is relevant, e.g., in systems with particle number conservation, but will be neglected here since we are only interested in scalings that are exponential in $L$.

While our parametrization of $\rho(L)$ discussed above is standard, the characterization of the function $|f_0|^2$ is more subtle. 
We parametrize the latter as
\begin{equation} \label{def_f0_parametrization}
    |f_0|^2 = |\overline{f_0(h,L)}|^2 \exp\{\xi/\eta_0\} \;.
\end{equation}
In Eq.~\eqref{def_f0_parametrization}, the term $|\overline{f_0(h,L)}|^2$ describes a smooth dependence on the model parameter $h$, and the dependence on $L$ that is at most polynomial in $L$.
The polynomial dependence on $L$ is known and was reported in several previous works~\cite{dalessio_kafri_16, luitz_barlev_16, schoenle_jansen_21, leblond_mallayya_19, brenes_leblond_20, brenes_goold_20, leblond_rigol_20, richter_dymarsky_20, zhang_vidmar_22, balachandran_santos_23, wang_zhu_24, patil_rigol_25}.
If the dependence of $|f_0|^2$ on $L$ is at most polynomial, which was, to our knowledge, the case in all previous works (except for fading ergodicity~\cite{kliczkowski_swietek_24}, to be discussed below), then $\xi$ in Eq.~\eqref{def_f0_parametrization} should not depend on $L$. 
Here, we go beyond this paradigm and we parametrize $\xi$ as
\begin{equation}
\xi=\xi(h,L)\,,
\end{equation}
i.e., we allow for $L$ dependence on $\xi$,
which gives rise to the possibility for an exponential scaling of $|f_0|^2$ with $L$. 

Summarizing this section, we express the dominant contribution to $\sigma^2$ in Eq.~\eqref{eq:diag_fluc_ETH} as
\begin{equation} \label{def_sigma2_xi}
    \sigma^2 \propto \exp\left\{-\frac{L-\xi(h,L)}{\eta_0}\right\} \,.
\end{equation}
We denote $\xi$ as the ergodization length, since only for $L \gg \xi$, the system exhibits conventional ETH.

In passing, we note that in systems exhibiting diffusive or subdiffusive transport, the ergodization length $\xi$ is a constant that does not depend on $L$, and hence these systems comply with the conventional ETH.
This is a consequence of the expected scaling $|f(\omega>\Gamma)|^2 \propto \omega^{-a}$~\cite{vidmar_21, sels_polkovnikov_21, sels_polkovnikov_23, capizzi_wang_25}, where $a \leq 1$ and $\Gamma$ is the Thouless energy that decreases at most polynomially with $L$, and $|f(\omega<\Gamma)|^2$ exhibits no dependence on $\omega$.
Then, $|f_0|^2$ also increases at most polynomially with $L$~\cite{dalessio_kafri_16, luitz_barlev_16,  brenes_leblond_20, richter_dymarsky_20, schoenle_jansen_21}, and hence $\xi$ does not depend on $L$.
On the other hand, in the case of Lorentzian $f$ function, $|f(\omega)|^2\propto \Gamma/(\omega^2+\Gamma^2)$, one cannot exclude the possibility that $\xi$ depends on $L$.
A known example where $\xi$ depends on $L$ is fading ergodicity~\cite{kliczkowski_swietek_24, swietek_kliczkowski_26}, which is a consequence of the exponential decrease of $\Gamma$ with $L$~\cite{suntajs_22}.

\section{Two-parameter ansatz for the ETH violation}

The central question is whether there exists a simple and meaningful parametrization of the ergodization length $\xi$.
Here, we introduce the two-parameter ansatz for $\xi$,
\begin{eqnarray}
    \xi(h,L) = \theta(h)L + \lambda(h)\,,  \label{eq:two_parameter_ansatz}
\end{eqnarray}
in which the parameters $\theta$ and $\lambda$ may depend on the model parameter $h$, but not on the system size $L$.
Plugging Eq.~\eqref{eq:two_parameter_ansatz} into Eq.~\eqref{def_sigma2_xi}, one can express the scaling of the variance of diagonal matrix elements as
\begin{eqnarray} \label{def_sigma2_final}
    \sigma^2 \propto \exp\left\{-\frac{[1-\theta(h)]L - \lambda(h)}{\eta_0}\right\}\,.
\end{eqnarray}
The limit $\theta=0$ corresponds to the conventional ETH.
More generally, the variance $\sigma^2$ may decrease, or remain constant, in the thermodynamic limit $L\to\infty$.
For structureless observables, ${\cal A}(\bar E)=0$, which are normalized by the Hilbert--Schmidt norm~\footnote{The square of the Hilbert--Schmidt norm is, for structureless observables, defined as $||\hat A||^2 = {\rm Tr}\{\hat A^2\}/{\cal D}$, see also Ref.~\cite{lydzba_swietek_24}. Normalized observables satisfy $||\hat A||=1$.}, the variance (calculated over the entire spectrum) satisfies $\sigma^2 \leq 1$.
It is then reasonable to bound the parameters $\theta$ and $\lambda$ for any normalized observable as $0\leq \theta \leq 1$ and $\lambda(h) \lesssim {\cal O}(L)$.
This gives rise to two scenarios for the ETH violation in a given quantum system, which are characterized by either $\theta\to 1$ or $\lambda \to {\cal O}(L)$.
We refer to these scenarios as fading ergodicity and trapped ergodicity, respectively, which will be separately discussed below.

\begin{figure}[!t]
\includegraphics[width=1.0\columnwidth]{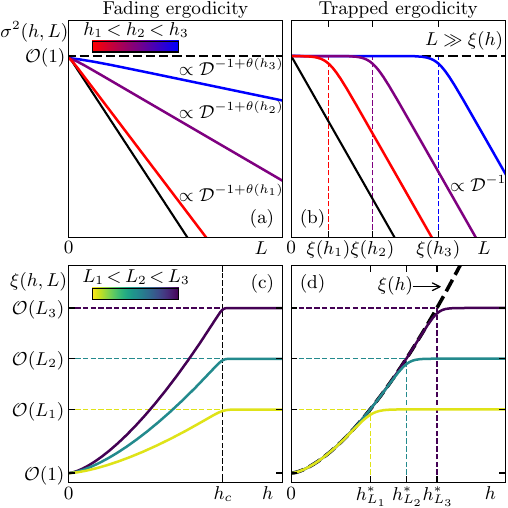}
\vspace{-0.2cm}
\caption{
Two scenarios for the ETH violation. (a)(c) fading ergodicity, and (b)(d) trapped ergodicity. 
(a)(b) The variance $\sigma^{2}$ as a function of system size $L$. The vertical axis is shown on a logarithmic scale. 
Black solid and dashed lines denote the conventional ETH and the complete breakdown of ETH, respectively. 
(c)(d) The ergodization length $\xi(h,L)$ as a function of model parameter $h$.
(c) $\xi$ increases with $L$, at fixed $h$, in the fading ergodicity regime at $h<h_c$.
(d) $\xi$ is independent of $L$, when $\xi(h)=\lambda(h)\lesssim {\cal O}(L)$, in the trapped ergodicity regime at $h<h_L^*$.
}
\label{fig:FE_vs_TE}
\end{figure}

\subsection{Fading ergodicity}

The first scenario of the ETH violation is the recently introduced fading ergodicity~\cite{kliczkowski_swietek_24}, which was shown to apply in the quantum sun model~\cite{kliczkowski_swietek_24, swietek_lydzba_25, swietek_hopjan_25}, as well as in structured random-matrix theory models such as the Rosenzweig--Porter model and the ultrametric model~\cite{swietek_kliczkowski_26}.
It is characterized by the parameter $\theta(h)$~\footnote{In the language of Ref.~\cite{kliczkowski_swietek_24} that introduced the parameter $\eta$ to characterize fading ergodicity, $1-\theta = 2/\eta$.} that increases towards 1 [see Fig.~\ref{fig:FE_vs_TE}(a)].
The ergodicity breaking critical point $h=h_c$, at which $\theta(h=h_c)=1$ and $\sigma^2={\cal O}(1)$, is characterized by the complete breakdown of the ETH, which corresponds to the horizontal dashed line in Fig.~\ref{fig:FE_vs_TE}(a).

Another manifestation of fading ergodicity is the $L$-dependent ergodization length $\xi$, see Fig.~\ref{fig:FE_vs_TE}(c), which diverges when $L\to\infty$ at fixed $h$.
Note that this occurs despite $\xi < L$ and consequently, the variance still vanishes, $\sigma^2 \to 0$ [see Eq.~\eqref{def_sigma2_xi}].

\subsection{Trapped ergodicity}

Another scenario of ETH violation is trapped ergodicity, which is the new result of this work.
It is characterized by $\theta(h) = 0$ and the parameter $\lambda(h) = \xi(h)$ that increases with $h$.
Hence, the "trapping'' is manifested as the shift on the horizontal axis of the decay of $\sigma^2$, with a decay rate at sufficiently large $L$ that is consistent with the conventional ETH, $\sigma^2 \propto \exp\{-L/\eta_0\} \approx {\cal D}^{-1}$, see the sketch in Fig.~\ref{fig:FE_vs_TE}(b).
A related manifestation of trapped ergodicity is the $L$-independent ergodization length, $\xi(h)$, as shown in Fig.~\ref{fig:FE_vs_TE}(d).

The subtle point about trapped ergodicity is that, as discussed below Eq.~\eqref{def_sigma2_final}, the ergodization length $\xi(h)$ should not exceed the system size, i.e., $\xi(h) \lesssim {\cal O}(L)$.
As a consequence, whenever $\xi(h)$ becomes larger than ${\cal O}(L)$, the numerical analysis in finite systems yields $\xi(h) \approx {\cal O}(L)$, and hence $\sigma^2 = {\cal O}(1)$.
This scenario is not very exotic, since one can easily imagine the ergodization length exceeding the largest system sizes available by exact diagonalization techniques, which yield $L_{\rm max} = {\cal O}(20)$.
Hence, it may be tempting to argue that $\sigma^2 = {\cal O}(1)$ in the thermodynamic limit and that the system exhibits a complete breakdown of the ETH.

A crossover from the conventional ETH-like scaling, $\sigma^2 \propto {\cal D}^{-1}$, to a complete breakdown of ETH, $\sigma^2 = {\cal O}(1)$, emerges at the parameter strength $h=h^*$ when $\xi(h)$ departs from the $L$-independent function.
This occurs at $\xi(h=h^*)= {\cal O}(L)$.
If $\xi(h)$ is a smooth function of $h$, e.g., it increases algebraically as $\xi(h) \propto h^\mu$, as shown in Fig.~\ref{fig:FE_vs_TE}(d), it follows that $h^*$ diverges with $L$, i.e., $h^* = h_L^*$, and no ergodicity breaking phase transition occurs in the thermodynamic limit.
Below we show two numerical examples of systems in which $\xi(h)$ increases algebraically with $h$ and is independent of $L$, as long as $\xi(h) \lesssim {\cal O}(L)$.
However, trapped ergodicity may not exclusively be associated with the absence of an ergodicity-breaking phase transition.
One may, e.g., imagine the divergence of the ergodization length as $\xi(h) \propto |h-h_c|^{-\nu}$, giving rise to the phase transition at $h=h_c$.
Still, we are currently not aware of physical examples that exhibit such scaling properties.


\begin{figure*}[!t]
\includegraphics[width=\textwidth]{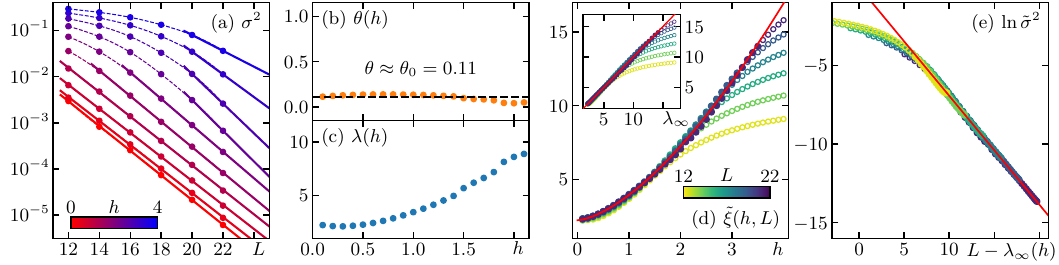}
\vspace{-0.2cm}
\caption{
Evidence of trapped ergodicity in the disordered $J_{1}$-$J_{2}$ model.
(a)~Variance $\sigma^2$ vs $L$ at various $h$.
The solid lines are linear fits to $\ln \sigma^2$ vs $L$ for the points at large $L$; see Appendix~\ref{app:numerics} for computational details.
The dashed curves are guides for the eye.
(b)(c)~The extracted parameters $\theta(h)$ and $\lambda(h)$ from the fits in (a).
The dashed line in (b) is the average $\theta_{0} = 0.11$ of $\theta(h)$ in the interval $h \in [0,2]$.
(d)~Scaled ergodization length $\tilde{\xi}(h,L) = \xi(h,L) - \theta_{0}L$ vs $h$ in the main panel.
The filled symbols correspond to the values of $h, L$ at which $r \approx r_{\rm GOE}$, while open symbols correspond to $r < r_{\rm GOE}$; see also Fig.~\ref{fig:gap_ratio} in Appendix~\ref{app:gapratio}.
The solid line is a fit of the function $\lambda_\infty(h) = a_0 h^{\mu} + a_{1}$ to the results for $\tilde{\xi}(h,L)$ in the $L$-independent regime.
We get $a_0 = 1.73$, $a_1 = 2.18$ and $\mu = 1.55$.
We show $\tilde{\xi}(h,L)$ vs $\lambda_\infty(h)$ in the inset. 
(e) Logarithm of the scaled variance $\tilde{\sigma}^2$ vs $L-\lambda_\infty(h)$; see the main text for details.
The solid line is the function $-(L-\lambda_{\infty})/\eta_{0}$, suggested by Eq.~\eqref{def_sigma2_final}.
}
\label{fig:num_ASD}
\end{figure*}

\section{Numerical examples}

As numerical examples, we study the $J_{1}$-$J_{2}$ Heisenberg chain \cite{suntajs_20_a, suntajs_20_b} with onsite potentials,
\begin{eqnarray}
\hspace*{-0.4cm}
\hat{H} = J_{1}\sum_{j=1}^{L-1}\hat{\bm{S}}_{j}\cdot \hat{\bm{S}}_{j+1} + J_{2}\sum_{j=1}^{L-2}\hat{\bm{S}}_{j}\cdot \hat{\bm{S}}_{j+2}  + \sum_{j=1}^{L}h_{j}\hat{S}^{z}_{j}, \label{eq:Hamiltonain}
\end{eqnarray}
where $\hat{\bm{S}}_{j} = (\hat{\sigma}^{x}_{j}, \hat{\sigma}^{y}_{j}, \hat{\sigma}^{z}_{j} )/2$ are spin-1/2 operators at site $j$. 
We set $J_1 =1$ and $J_2 = (\sqrt{5}-1)/2 \approx 0.618$, and we consider even system sizes $L \in [12, 22]$ using polynomially filtered exact diagonalization~\cite{sierant_lewenstein_20, pintar_pawlik_26}.

In the main text, we study the $J_1$-$J_2$ model with onsite disorder $h_{j} = h\epsilon_{j}$, where $h$ is the disorder strength, and the random numbers $\epsilon_{j}$ are drawn from a uniform distribution $[-1, 1]$.
An analogous study is carried out in Appendix~\ref{app:stark} for the Stark $J_1$-$J_2$ model that includes a linear potential.
In both cases, we numerically compute the variance $\sigma^2$, see Eq.~(\ref{eq:def_diag_fluc}), for the spin operator $\hat{S}_{L/2}^{z}$ at the central lattice site $j=L/2$.

Figure~\ref{fig:num_ASD}(a) shows $\sigma^2$ versus $L$ at various $h$. 
In the weakly disordered regime $h\lesssim 1.5$ (red dots), the decay of $\sigma^2$ is consistent with the conventional ETH, $\sigma^2 \propto \exp\{-L/\eta_0\}\approx {\cal D}^{-1}$, for all $L$ under consideration. 
At larger disorder, we observe horizontal shifts of the decays of $\sigma^2$, without a considerable change in the slopes.
This is an indication of trapped ergodicity.

We next extract the parameters $\theta(h)$ and $\lambda(h)$ from Eq.~\eqref{def_sigma2_final} assuming equality instead of proportionality; see also Appendix~\ref{app:numerics} for the details of the fitting procedure.
Results of this quantitative analysis are shown in Figs.~\ref{fig:num_ASD}(b) and~\ref{fig:num_ASD}(c), respectively.
We observe that $\theta(h)$ is roughly a constant that is very close to zero (we fit $\theta$ by a constant $\theta_0$ for $h\in[0,2]$ and get $\theta_0=0.11$), while $\lambda(h)$ is an increasing function of $h$. 
This behavior is a clear signature of trapped ergodicity, and it is very different from fading ergodicity, in which $\theta(h)$ increases with $h$~\cite{kliczkowski_swietek_24, swietek_kliczkowski_26}. 
We note that small nonzero values of $\theta$ in Fig.~\ref{fig:num_ASD}(b) may originate from polynomial corrections to the ansatz in Eq.~\eqref{def_sigma2_final}, and hence they are not inconsistent with $\theta=0$ in the thermodynamic limit.

A complementary approach to the fitting procedure that yields $\theta(h)$ and $\lambda(h)$ is to directly extract the ergodization length $\xi(h,L)$, assuming equality instead of proportionality in Eq.~\eqref{def_sigma2_xi}, i.e., to calculate $\xi(h,L) = \eta_0 \ln\sigma^2 + L$.
Small finite-size effects can be removed if one also takes into account the small nonzero value of $\theta(h)=\theta_0$, and studies the scaled ergodization length
$\tilde{\xi}(h,L) = \xi(h,L) - \theta_{0}L$.

Results for $\tilde{\xi}(h,L)$ versus $h$ are shown in Fig.~\ref{fig:num_ASD}(d) at various $L$.
Most importantly, at small $h$, $\tilde{\xi}(h,L)$ is independent of $L$, and it follows from Eq.~\eqref{eq:two_parameter_ansatz} that in this regime, $\tilde{\xi}(h,L) \equiv \lambda_\infty(h)$.
The results in this regime are well fitted by the asymptotic function 
$\lambda_\infty(h) = a_0 h^{\mu} + a_{1}$,
where $a_0, a_1$ and $\mu$ are fitting parameters.

It is remarkable that the $L$-independent regime of $\tilde{\xi}(h,L)$ almost exactly coincides with the departure of the nearest level spacing ratio $r$ from the Gaussian orthogonal ensemble (GOE) prediction $r_{\rm GOE}$; see Fig.~\ref{fig:gap_ratio} in Appendix~\ref{app:gapratio} for the results for $r$ vs $h$ at various $L$.
In Fig.~\ref{fig:num_ASD}(d), filled symbols correspond to the values of $h$ and $L$ when $r \approx r_{\rm GOE}$, and they all lie on the $L$ independent curve described by $\lambda_\infty(h)$.
On the other hand, open symbols correspond to the values of $h$ and $L$ when $r < r_{\rm GOE}$, and in this regime the $L$-dependence of $\tilde{\xi}(h,L)$ becomes important.
At larger $h$, $\tilde{\xi}(h,L)$ eventually saturate at values given by $\tilde{\xi}(h,L) \approx \mathcal{O}(L)$.

Our analysis of trapped ergodicity identifies the relevant variable for the scaling of variance to be $L-\lambda_\infty(h)$.
This can be tested numerically by plotting $\ln \sigma^2$ vs $L-\lambda_\infty(h)$.
As in Fig.~\ref{fig:num_ASD}(d), we minimize finite-size effects by studying the scaled variance $\ln \tilde{\sigma}^{2} = \ln \sigma^{2} - \theta_{0}L/\eta_{0}$.
Results for $\ln \tilde{\sigma}^{2}$ are shown in Fig.~\ref{fig:num_ASD}(e) at various $h$ and $L$.
They exhibit perfect scaling collapse in the regime where $\ln \tilde{\sigma}^{2} \propto L-\lambda_\infty(h)$, and a reasonably good scaling collapse even in the regime where $L-\lambda_\infty(h) \approx 0$.

We note that a large, but $L$-independent ergodization length $\lambda_\infty(h)$, which emerges in the scaling of observable matrix elements studied here, may be consistent with an $L$-independent correction to the eigenstate participation entropies, observed in a study of the disordered XXZ model~\cite{mace_19}.
Still, our results for $\lambda_\infty$ vs $h$ in Fig.~\ref{fig:num_ASD}(d) do not exhibit a divergence at finite $h$, and hence do not imply emergence of an ergodicity breaking phase transition.

Finally, a study analogous to the one in Fig.~\ref{fig:num_ASD} is shown for the Stark $J_1$-$J_2$ model in Fig.~\ref{fig:num_stark} of Appendix~\ref{app:stark}.
Even though the results in Fig.~\ref{fig:num_stark} do not exhibit as remarkable scaling properties as those in Fig.~\ref{fig:num_ASD}, we still argue that the ETH violation in the Stark $J_1$-$J_2$ model can be reasonably well described using the framework of trapped ergodicity.

\section{Conclusions}

The main result of this work is the introduction of a common theoretical framework to describe two distinct scenarios of the ETH breakdown.
While the first scenario, fading ergodicity, has been established before~\cite{kliczkowski_swietek_24, swietek_lydzba_25, swietek_hopjan_25, swietek_kliczkowski_26}, we here showed that the second scenario, dubbed trapped ergodicity, applies to widely studied models such as the disordered $J_1$-$J_2$ spin-1/2 chain.
Our results not only offer a perspective into the analysis of finite-size effects at the boundary of ergodicity, but also open the door to establishing firm connections between quantum thermalization and ergodicity breaking transitions.

While our analysis focused on the diagonal matrix elements, it would be desirable to carry out a complementary study for the off-diagonal matrix elements.
This has been so far studied only in the fading ergodicity regime, where similarity between the scaling properties of the diagonal and off-diagonal matrix elements has been observed, provided that the energy difference $\omega$ of the latter is lower than the Thouless energy~\cite{kliczkowski_swietek_24, swietek_kliczkowski_26}.
Another immediate question opened by this work is to find appropriate toy models that would allow for a deeper analytical insight into the scaling properties of matrix elements. 
While for fading ergodicity, two such models have been recently identified in the context of structured random-matrix models, namely, the Rosenzweig--Porter model and the ultrametric model~\cite{swietek_kliczkowski_26}, similar models of trapped ergodicity are still missing.

\acknowledgements 

We acknowledge support from the Slovenian Research and Innovation Agency (ARIS), Research core funding Grants No.~P1-0044, No.~N1-0273 and No.~J1-50005, as well as the Consolidator Grant No. Boundary-101126364 of the European Research Council (ERC).
We gratefully acknowledge the High Performance Computing Research Infrastructure Eastern Region (HCP RIVR) consortium~\cite{vega1} and European High Performance Computing Joint Undertaking (EuroHPC JU)~\cite{vega2} for funding this research by providing computing resources of the HPC system Vega at the Institute of Information sciences~\cite{vega3}.

\section*{Data Availability}
Research data associated with this article are available on Zenodo~\cite{ohgane_2026_23983056}.


\appendix

\section{Details of the numerical calculations} \label{app:numerics}

We calculate the eigenstates $\ket{m}$ at eigenenergies $E_{m}$ of the Hamiltonian $\hat{H}$ in Eq.~(\ref{eq:Hamiltonain}) using the polynomially filtered exact diagonalization (POLFED) at $L=14 - 22$~\cite{sierant_lewenstein_20, pintar_pawlik_26}, and full exact diagonalization at $L=12$. 
We obtain the diagonal matrix elements $A_{mm} = \bra{m}\hat{A}\ket{m}$ of the observable $\hat{A} = \hat{S}^{z}_{L/2}$, which is normalized such that the Hilbert--Schmidt norm within the targeted symmetry sector is $|| \hat{A}||^{2}_{\mathrm{HS}} := \mathrm{Tr}\{\hat{A}^{2}\}/{\cal D} - (\mathrm{Tr}\{\hat{A}\}/{\cal D})^{2} = 1$~\cite{lydzba_swietek_24}.
We then calculate the variance $\sigma^2$ from Eq.~(\ref{eq:def_diag_fluc}), setting the width of the moving average to $M=30$.
We consider the energy window at the central energy $E_0 = \mathrm{Tr}\{\hat{H}\}/{\cal D}$ with the width $\delta E = 0.02\times \Delta E$, where $\Delta E ^{2} =\mathrm{Tr}\{\hat{H}^{2}\}/{\cal D} - E_0^{2}$.
If the energy window $\delta E$ requires calculation of more than 1500 eigenstates at $L\geq 14$, we consider at most 1500 eigenstates closest to $E_0$.
Finally, we average $\sigma^2$ over $N_{\mathrm{r}}$ disorder realizations, where $N_{\mathrm{r}}$ varies from $1.5\times 10^{2}$ to $2\times10^{4}$, depending on $L$.

After obtaining the variance $\sigma^2$, we extract the parameters $\theta(h)$ and $\lambda(h)$ in the two-parameter ansatz of Eq.~(\ref{eq:two_parameter_ansatz}) by fitting the data points as a function of $L$ at fixed $h$, using equality instead of proportionality in Eq.~\eqref{def_sigma2_final}.
In particular, a linear fit to $\ln \sigma^2$ yields the slope $a(h)$ and the intercept $b(h)$ according to $\ln \sigma^2 = a(h)L+b(h)$, and the parameters $\theta(h)$ and $\lambda(h)$ are given by the fitting parameters $a(h)$ and $b(h)$ as 
\begin{eqnarray}
    \theta(h) = 1 - \eta_{0}a(h), \quad \lambda(h) = \eta_{0}b(h). 
\end{eqnarray}
The parameter $\eta_{0}$ is $\eta_0 = 1/\ln(2) \approx 1.44$, which comes from the asymptotic exponential increase of the Hilbert space dimension, ${\cal D} = L!/(L/2)!(L/2)! \asymp 2^L$, in the zero-magnetization sector, $\sum_{j=1}^{L}\hat{S}^{z}_{j}=0$.

Red dots in Fig.~\ref{fig:num_ASD}(a) of the main text show that the linear fit to $\ln \sigma^2$ works well for sufficiently small $h$. In contrast, for larger $h$, see the purple dots in Fig.~\ref{fig:num_ASD}(a), $\ln \sigma^2$ is not well described by a linear function over the full range of numerically accessible $L$.
Still, at sufficiently large $L$, $\ln \sigma^2$ remains well described by a linear function. 

We restrict the range of $L$ used for the linear fit as follows. 
We denote the set of values of $L$ included in the fit as $\Omega_n$, where $\Omega_0 = \{12, 14, 16, 18, 20, 22\}$, $\Omega_1$ includes all the values from $\Omega_0$ except for $L=12$, etc.
We define the fitting error,
\begin{eqnarray}
    \chi_n^2 = \frac{1}{|\Omega_n|}\sum_{L\in \Omega_n}|aL+b-\ln\sigma(L)|^{2}, \label{eq:fitting_error}
\end{eqnarray}
where $a$ and $b$ are fitting parameters obtained from the linear fit to the numerical results for $\ln \sigma^2(L)$ in the fitting range $\Omega_n$, and $|\Omega_n|$ denotes the number of the elements in $\Omega_n$. 
In the numerical procedure, we first set $n=0$, perform a linear fit to $\ln \sigma^2$, and compute the fitting error $\chi_0^2$ from Eq.~(\ref{eq:fitting_error}).
If $\chi_0^2$ is below the threshold $\chi^{2}_{\mathrm{thr}}$, the procedure is terminated.
Otherwise, we increase $n$ to $n+1$ and repeat the fit.
The procedure is terminated either when $\chi_n^2 < \chi_{\rm thr}^2$, or when the number of elements in $\Omega_n$ drops to 2.
The results for $\theta(h)$ and $\lambda(h)$ are shown in Figs.~\ref{fig:num_ASD}(b) and~\ref{fig:num_ASD}(c).

\begin{figure}[!t]
\includegraphics[width=1.0\columnwidth]{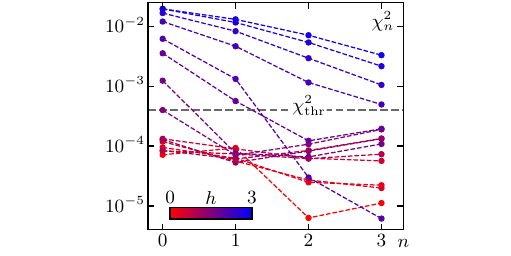}
\vspace{-0.2cm}
\caption{
    Fitting error $\chi^2_{n}$, see Eq.~(\ref{eq:fitting_error}), vs $n$. The horizontal dashed line is the threshold $\chi^2_{\mathrm{thr}} = 4\times 10^{-4}$. The thin dashed lines between the adjacent dots are guides of the eye.
}
\label{fig:fitting_error}
\end{figure}
 
We set the threshold $\chi^2_{\mathrm{thr}} = 4\times 10^{-4}$ in the disordered $J_1$-$J_2$ model. 
We choose this value so that in the conventional ETH regime, i.e., at $h \approx 1$, the typical values of $\chi_0^2$ are below $\chi^2_{\mathrm{thr}}$.
Moreover, the values of $\chi_n^2$ in the conventional ETH regime should be roughly independent of $n$.
Results in Fig.~\ref{fig:fitting_error} suggest that the latter is indeed the case, and that our choice of $\chi^2_{\mathrm{thr}}$ is reasonable.

\section{Gap ratio} \label{app:gapratio}

In Figs.~\ref{fig:num_ASD}(d) and~\ref{fig:num_ASD}(e) in the main text, filled symbols correspond to the results in the regime of $h$ and $L$ when the gap ratio statistics match the GOE predictions.
Here, we explain how we define this regime.

The gap ratio is defined for the adjacent eigenenergies $E_{m-1}, E_{m}, E_{m+1}$ as 
$r_{m} = \frac{\min(s_{m}, s_{m+1})}{\max(s_{m}, s_{m+1})}$, 
where $s_m$ is the energy gap, $s_{m} = E_{m}-E_{m-1}$. 
We first average $r_m$ over the same energy window as in the calculation of the variance $\sigma^2$, and then over different Hamiltonian realizations to obtain $r$. 
We then introduce the rescaled gap ratio,
\begin{eqnarray}
    \tilde{r} := \frac{r-r_{\mathrm{P}}}{r_{\mathrm{GOE}}-r_{\mathrm{P}}}\,,
    \label{eq:ave_res_gap_ratio}
\end{eqnarray}
where $r_{\mathrm{GOE}} = 0.5307$ is the GOE prediction and $r_{\mathrm{P}} = 0.3863$ is the prediction of the Poisson statistics, such that one expects $\tilde{r} \in [0,1]$.

\begin{figure}[!t]
\includegraphics[width=1.0\columnwidth]{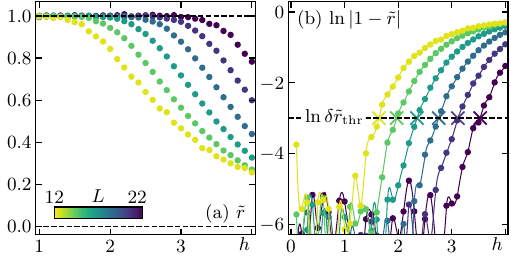}
\vspace{-0.2cm}
\caption{
    Rescaled gap ratio $\tilde{r}$, see Eq.~(\ref{eq:ave_res_gap_ratio}), in the disordered $J_{1}$-$J_{2}$ model.
    (a)~$\tilde{r}$ vs $h$ at various $L$. The upper and lower dashed lines represent the GOE and Poisson values, respectively. 
    (b)~$\ln|1-\tilde{r}|$ vs $h$ at various $L$.
    The horizontal dashed line shows the threshold $\ln \delta \tilde{r}_{\rm thr} = -3$. The solid curves represent the interpolations of the data points. 
    The crosses mark the points at which the interpolating curves of $\ln|1-\tilde{r}|$ cross to $\ln \delta \tilde{r}_{\rm thr}$. 
}
\label{fig:gap_ratio}
\end{figure}

Results for $\tilde{r}$ vs $h$ at various $L$ in the disordered $J_1$-$J_2$ model are shown in Fig.~\ref{fig:gap_ratio}(a). 
We denote by $h_{r}^{*}(L)$ the disorder strength at which $\tilde{r}$ starts to deviate from $\tilde{r}=1$, which shifts to larger values as $L$ increases. 
Specifically, we define $h^{*}_{r}(L)$ as the value of $h$ at which $|1-\tilde{r}|$ reaches the threshold value $\delta \tilde{r}_{\rm thr}$. 
Figure~\ref{fig:gap_ratio}(b) shows $\ln|1-\tilde{r}|$, where the dashed line represents $\ln\delta\tilde{r}_{\rm thr}$. 
We choose the value of $\delta \tilde{r}_{\rm thr}= \mathrm{e}^{-3} \approx 0.05$, such that it lies above the numerical noise, which occurs at $\ln|1-\tilde r| \lesssim -4$. 
The value of $h$ at which $\ln|1- \tilde{r}|$ crosses $\ln\delta\tilde{r}_{\rm thr}$ is estimated using an interpolating cubic spline implemented in SciPy and is marked by crosses in Fig.~\ref{fig:gap_ratio}(b).

\begin{figure*}[!t]
\includegraphics[width=\textwidth]{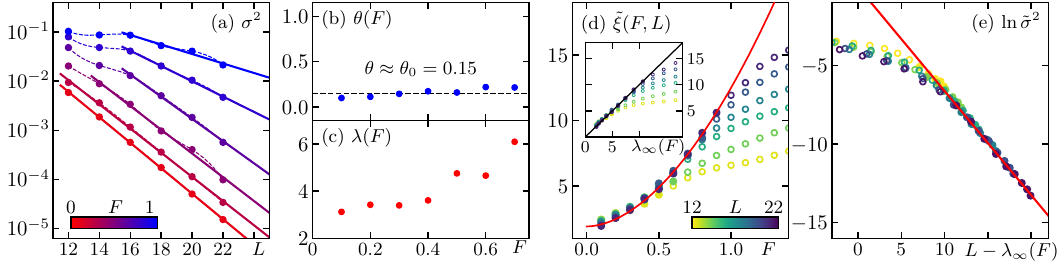}
\vspace{-0.2cm}
\caption{
Evidence of trapped ergodicity in the Stark $J_{1}$-$J_{2}$ model.
(a)~Variance $\sigma^2$ vs $L$ at various $F$.
The solid lines are linear fits to $\ln \sigma^2$ vs $L$ for the points at large $L$.
The dashed curves are guides for the eye.
(b)(c)~The extracted parameters $\theta(F)$ and $\lambda(F)$ from the fits in (a).
The dashed line in (b) is the average $\theta_0=0.15$ of $\theta(F)$ in the interval $F \in [0, 0.7]$.
(d)~Scaled ergodization length $\tilde{\xi}(F,L) = \xi(F,L) - \theta_{0}L$ vs $F$ in the main panel.
The filled symbols correspond to the values of $F$ and $L$ at which $r \approx r_{\rm GOE}$, while open symbols correspond to $r < r_{\rm GOE}$.
The solid line is a fit of the function $\lambda_\infty(F) = a_0 F^{\mu} + a_{1}$ to the results for $\tilde{\xi}(F,L)$ in the $L$-independent regime.
We get $a_0 = 10.58$, $a_1 = 1.90$ and $\mu=1.83$.
We show $\tilde{\xi}(F,L)$ vs $\lambda_\infty(F)$ in the inset. 
(e) Logarithm of the scaled variance $\tilde{\sigma}^2$ vs $L-\lambda_\infty(F)$.
The solid line is the function $-(L-\lambda)/\eta_{0}$, suggested by Eq.~\eqref{def_sigma2_final}.
}
\label{fig:num_stark}
\end{figure*}

\section{Result for the Stark $J_1$-$J_2$ model} \label{app:stark}

Here, we demonstrate trapped ergodicity in the Stark $J_{1}$-$J_{2}$ model [see Eq.~(\ref{eq:Hamiltonain})]. 
The Stark potential is defined as $h_{j} = F\cdot(j-j_{\mathrm{c}}) + 0.5\cdot\epsilon_{j}$, where $F$ is the potential strength, $j_{\mathrm{c}} = (L+1)/2$ is the central position of the chain, and $\epsilon_{j}$ are random numbers that satisfy the antisymmetrized condition $\epsilon_{j}=-\epsilon_{L-j+1}$. 
Models similar to the one considered here were proposed to exhibit ergodicity breaking phenomena~\cite{nieuwenburg_baum_19, schulz_hooley_19}.

The Hamiltonian conserves not only the total magnetization $\hat{S}^{z}_{\mathrm{tot}} = \sum_{j=1}^{L}\hat{S}^{z}_{j}$, but also the parity $\hat{P}$ originating from the symmetry of the potential $h_{j}$. The parity operator is given by a product of two operators, i.e., $\hat{P} = \hat{P}_{\mathrm{flip}} \hat{P}_{\mathrm{reflect}}$, where 
$\hat{P}_{\mathrm{flip}} = \prod_{j=1}^{L}\hat{\sigma}_{j}^{x}$ and $\hat{P}_{\mathrm{reflect}} = \prod_{j=1}^{L/2}\frac{1+\hat{\bm{\sigma}}_{j}\cdot \hat{\bm{\sigma}}_{L-j+1}}{2}$.
Since $\hat{P}^{2} = 1$, the eigenvalues of $\hat{P}$ are $\pm 1$. $\hat{P}$ changes the sign of the total magnetization, $\hat{P}\hat{S}^{z}_{\mathrm{tot}}\hat{P} = -\hat{S}^{z}_{\mathrm{tot}}$. 
Thus, $\hat{S}^{z}_{\mathrm{tot}}$ and $\hat{P}$ commute and have simultaneous eigenstates only within the sector $\hat{S}^{z}_{\mathrm{tot}}=0$. We restrict our calculation to the symmetry sector labeled by $\hat{S}_{\mathrm{tot}}^z = 0$ and $\hat{P} = +1$, for which the Hilbert-space dimension is ${\cal D} = \frac{1}{2}\bigl( \frac{L!}{(L/2)!(L/2)!} + 2^{L/2} \bigr)$.
For the system sizes under consideration, $L \leq 22$, we parametrize ${\cal D}$ by the function $\mathrm{const} \times e^{L/\eta_{0}}$, with $\eta_{0} \approx 1.509$ obtained from a fit to the numerical values of $\cal D$ in the range $L \in [10,30]$.

Figure~\ref{fig:num_stark}(a) shows the variance $\sigma^2$ from Eq.~\eqref{eq:def_diag_fluc}. 
At weak potential, $F\lesssim 0.3$, the decay of $\sigma^2$ is consistent with the conventional ETH, $\sigma^2 \propto \exp\{-L/\eta_{0}\} \approx \mathcal{D}^{-1}$, for all $L$ under consideration. At larger $F$ (but still $F \lesssim 0.7$), we observe horizontal shifts of the decays of $\sigma^2$, without a considerable change in slopes. 

The extracted parameters $\theta(F)$ and $\lambda(F)$ from Eq.~(\ref{eq:two_parameter_ansatz}) are shown in Figs.~\ref{fig:num_stark}(b) and~\ref{fig:num_stark}(c), respectively. 
The fitting procedure to extract $\theta(F)$ and $\lambda(F)$ is analogous to the one for the disordered $J_1$-$J_2$ model, and we set the fitting-error threshold (see Appendix~\ref{app:numerics}) to $\chi^2_{\mathrm{thr}} = 2\times 10^{-2}$.
We observe that $\theta(F)$ is roughly a constant that is  close to zero, while $\lambda(F)$ is an increasing function of $F$.
This behavior is a signature of trapped ergodicity. 

We next consider the ergodization length, defined as $\xi(F,L) = \eta_{0}\ln \sigma^2 + L$, using Eq.~(\ref{def_sigma2_xi}). 
As in Fig.~\ref{fig:num_ASD}(d), we minimize small finite-size effects by studying the rescaled ergodization length $\tilde{\xi}(F, L) = \xi(F,L) - \theta_{0}L$. 
Results for $\tilde{\xi}(F,L)$ vs $F$ are shown in Fig.~\ref{fig:num_stark}(d) at various $L$. 
At small $F$, $\tilde{\xi}(F,L)$ is an $L$-independent function. 
The results in this regime are well fitted by the asymptotic function $\lambda_{\infty}(F) = a_{0}F^{\mu} + a_{1}$, where $a_{0}, a_{1}$ and $\mu$ are fitting parameters.
The $L$-independent regime of $\tilde{\xi}(F,L)$ coincides with the regime in which results for $\tilde{\xi}(F,L)$ are represented by filled symbols.
The latter corresponds to the values of $F$ and $L$ for which $r \approx r_{\mathrm{GOE}}$ (we determine this regime by the same method as in Fig.~\ref{fig:gap_ratio}(b) for the disordered $J_1$-$J_2$ model). 

Finally, we test the suggested scaling from the main text using the scaling variable $L-\lambda_{\infty}(F)$.
Figure~\ref{fig:num_stark}(e) shows $\ln \tilde{\sigma}^2$ versus $L-\lambda_{\infty}(F)$, where finite-size effects are minimized by defining $\ln \tilde \sigma^2  = \ln \sigma^2 - \theta_0 L/\eta_{0}$. 
The results exhibit a scaling collapse in the regime when $\ln \tilde{\sigma}^2 \propto L - \lambda_{\infty}(F)$, which suggests the relevance of trapped ergodicity for describing the ETH violation in the Stark $J_1$-$J_2$ model.

\bibliographystyle{biblev1}
\bibliography{references}

\begin{thebibliography}{10}
\expandafter\ifx\csname url\endcsname\relax
  \def\url#1{{\tt #1}}\fi
\expandafter\ifx\csname urlprefix\endcsname\relax\def\urlprefix{URL }\fi
\expandafter\ifx\csname bibinfo\endcsname\relax\def\bibinfo#1#2{#2}\fi
\expandafter\ifx\csname eprint\endcsname\relax\def\eprint#1{\url{#1}}\fi

\bibitem{deutsch_91}
\bibinfo{author}{J.~M. Deutsch}, \bibinfo{title}{Quantum statistical mechanics in a closed system}, \bibinfo{journal}{\href{http://dx.doi.org/10.1103/PhysRevA.43.2046}{Phys. Rev. A}} \href{http://dx.doi.org/10.1103/PhysRevA.43.2046}{{\bf \bibinfo{volume}{43}}, \bibinfo{pages}{2046}}  (\href{http://dx.doi.org/10.1103/PhysRevA.43.2046}{\bibinfo{year}{1991}}).

\bibitem{srednicki_94}
\bibinfo{author}{M.~Srednicki}, \bibinfo{title}{Chaos and quantum thermalization}, \bibinfo{journal}{\href{http://dx.doi.org/10.1103/PhysRevE.50.888}{Phys. Rev. E}} \href{http://dx.doi.org/10.1103/PhysRevE.50.888}{{\bf \bibinfo{volume}{50}}, \bibinfo{pages}{888}}  (\href{http://dx.doi.org/10.1103/PhysRevE.50.888}{\bibinfo{year}{1994}}).

\bibitem{dalessio_kafri_16}
\bibinfo{author}{L.~D'Alessio}, \bibinfo{author}{Y.~Kafri}, \bibinfo{author}{A.~Polkovnikov}, and \bibinfo{author}{M.~Rigol}, \bibinfo{title}{From quantum chaos and eigenstate thermalization to statistical mechanics and thermodynamics}, \bibinfo{journal}{\href{http://dx.doi.org/10.1080/00018732.2016.1198134}{Adv. Phys.}} \href{http://dx.doi.org/10.1080/00018732.2016.1198134}{{\bf \bibinfo{volume}{65}}, \bibinfo{pages}{239}}  (\href{http://dx.doi.org/10.1080/00018732.2016.1198134}{\bibinfo{year}{2016}}).

\bibitem{rigol_dunjko_08}
\bibinfo{author}{M.~Rigol}, \bibinfo{author}{V.~Dunjko}, and \bibinfo{author}{M.~Olshanii}, \bibinfo{title}{Thermalization and its mechanism for generic isolated quantum systems}, \bibinfo{journal}{\href{http://dx.doi.org/10.1038/nature06838}{Nature (London)}} \href{http://dx.doi.org/10.1038/nature06838}{{\bf \bibinfo{volume}{452}}, \bibinfo{pages}{854}}  (\href{http://dx.doi.org/10.1038/nature06838}{\bibinfo{year}{2008}}).

\bibitem{rigol09}
\bibinfo{author}{M.~Rigol}, \bibinfo{title}{Quantum quenches and thermalization in one-dimensional fermionic systems}, \bibinfo{journal}{\href{http://dx.doi.org/10.1103/PhysRevA.80.053607}{Phys. Rev. A}} \href{http://dx.doi.org/10.1103/PhysRevA.80.053607}{{\bf \bibinfo{volume}{80}}, \bibinfo{pages}{053607}}  (\href{http://dx.doi.org/10.1103/PhysRevA.80.053607}{\bibinfo{year}{2009}}).

\bibitem{rigol_09a}
\bibinfo{author}{M.~Rigol}, \bibinfo{title}{Breakdown of thermalization in finite one-dimensional systems}, \bibinfo{journal}{\href{http://dx.doi.org/10.1103/PhysRevLett.103.100403}{Phys. Rev. Lett.}} \href{http://dx.doi.org/10.1103/PhysRevLett.103.100403}{{\bf \bibinfo{volume}{103}}, \bibinfo{pages}{100403}}  (\href{http://dx.doi.org/10.1103/PhysRevLett.103.100403}{\bibinfo{year}{2009}}).

\bibitem{santos_rigol_10b}
\bibinfo{author}{L.~F. Santos} and \bibinfo{author}{M.~Rigol}, \bibinfo{title}{Localization and the effects of symmetries in the thermalization properties of one-dimensional quantum systems}, \bibinfo{journal}{\href{http://dx.doi.org/10.1103/PhysRevE.82.031130}{Phys. Rev. E}} \href{http://dx.doi.org/10.1103/PhysRevE.82.031130}{{\bf \bibinfo{volume}{82}}, \bibinfo{pages}{031130}}  (\href{http://dx.doi.org/10.1103/PhysRevE.82.031130}{\bibinfo{year}{2010}}).

\bibitem{rigol_santos_10}
\bibinfo{author}{M.~Rigol} and \bibinfo{author}{L.~F. Santos}, \bibinfo{title}{Quantum chaos and thermalization in gapped systems}, \bibinfo{journal}{\href{http://dx.doi.org/10.1103/PhysRevA.82.011604}{Phys. Rev. A}} \href{http://dx.doi.org/10.1103/PhysRevA.82.011604}{{\bf \bibinfo{volume}{82}}, \bibinfo{pages}{011604}}  (\href{http://dx.doi.org/10.1103/PhysRevA.82.011604}{\bibinfo{year}{2010}}).

\bibitem{steinigeweg_herbrych_13}
\bibinfo{author}{R.~Steinigeweg}, \bibinfo{author}{J.~Herbrych}, and \bibinfo{author}{P.~Prelov\ifmmode~\check{s}\else \v{s}\fi{}ek}, \bibinfo{title}{Eigenstate thermalization within isolated spin-chain systems}, \bibinfo{journal}{\href{http://dx.doi.org/10.1103/PhysRevE.87.012118}{Phys. Rev. E}} \href{http://dx.doi.org/10.1103/PhysRevE.87.012118}{{\bf \bibinfo{volume}{87}}, \bibinfo{pages}{012118}}  (\href{http://dx.doi.org/10.1103/PhysRevE.87.012118}{\bibinfo{year}{2013}}).

\bibitem{khatami_pupillo_13}
\bibinfo{author}{E.~Khatami}, \bibinfo{author}{G.~Pupillo}, \bibinfo{author}{M.~Srednicki}, and \bibinfo{author}{M.~Rigol}, \bibinfo{title}{Fluctuation-dissipation theorem in an isolated system of quantum dipolar bosons after a quench}, \bibinfo{journal}{\href{http://dx.doi.org/10.1103/PhysRevLett.111.050403}{Phys. Rev. Lett.}} \href{http://dx.doi.org/10.1103/PhysRevLett.111.050403}{{\bf \bibinfo{volume}{111}}, \bibinfo{pages}{050403}}  (\href{http://dx.doi.org/10.1103/PhysRevLett.111.050403}{\bibinfo{year}{2013}}).

\bibitem{torresherrera_santos_14}
\bibinfo{author}{E.~J. Torres-Herrera} and \bibinfo{author}{L.~F. Santos}, \bibinfo{title}{Local quenches with global effects in interacting quantum systems}, \bibinfo{journal}{\href{http://dx.doi.org/10.1103/PhysRevE.89.062110}{Phys. Rev. E}} \href{http://dx.doi.org/10.1103/PhysRevE.89.062110}{{\bf \bibinfo{volume}{89}}, \bibinfo{pages}{062110}}  (\href{http://dx.doi.org/10.1103/PhysRevE.89.062110}{\bibinfo{year}{2014}}).

\bibitem{beugeling_moessner_14}
\bibinfo{author}{W.~Beugeling}, \bibinfo{author}{R.~Moessner}, and \bibinfo{author}{M.~Haque}, \bibinfo{title}{Finite-size scaling of eigenstate thermalization}, \bibinfo{journal}{\href{http://dx.doi.org/10.1103/PhysRevE.89.042112}{Phys. Rev. E}} \href{http://dx.doi.org/10.1103/PhysRevE.89.042112}{{\bf \bibinfo{volume}{89}}, \bibinfo{pages}{042112}}  (\href{http://dx.doi.org/10.1103/PhysRevE.89.042112}{\bibinfo{year}{2014}}).

\bibitem{sorg14}
\bibinfo{author}{S.~Sorg}, \bibinfo{author}{L.~Vidmar}, \bibinfo{author}{L.~Pollet}, and \bibinfo{author}{F.~Heidrich-Meisner}, \bibinfo{title}{{Relaxation and thermalization in the one-dimensional Bose-Hubbard model: A case study for the interaction quantum quench from the atomic limit}}, \bibinfo{journal}{\href{http://dx.doi.org/10.1103/PhysRevA.90.033606}{Phys. Rev. A}} \href{http://dx.doi.org/10.1103/PhysRevA.90.033606}{{\bf \bibinfo{volume}{90}}, \bibinfo{pages}{033606}}  (\href{http://dx.doi.org/10.1103/PhysRevA.90.033606}{\bibinfo{year}{2014}}).

\bibitem{steinigeweg_khodja_14}
\bibinfo{author}{R.~Steinigeweg}, \bibinfo{author}{A.~Khodja}, \bibinfo{author}{H.~Niemeyer}, \bibinfo{author}{C.~Gogolin}, and \bibinfo{author}{J.~Gemmer}, \bibinfo{title}{Pushing the limits of the eigenstate thermalization hypothesis towards mesoscopic quantum systems}, \bibinfo{journal}{\href{http://dx.doi.org/10.1103/PhysRevLett.112.130403}{Phys. Rev. Lett.}} \href{http://dx.doi.org/10.1103/PhysRevLett.112.130403}{{\bf \bibinfo{volume}{112}}, \bibinfo{pages}{130403}}  (\href{http://dx.doi.org/10.1103/PhysRevLett.112.130403}{\bibinfo{year}{2014}}).

\bibitem{kim_ikeda_14}
\bibinfo{author}{H.~Kim}, \bibinfo{author}{T.~N. Ikeda}, and \bibinfo{author}{D.~A. Huse}, \bibinfo{title}{Testing whether all eigenstates obey the eigenstate thermalization hypothesis}, \bibinfo{journal}{\href{http://dx.doi.org/10.1103/PhysRevE.90.052105}{Phys. Rev. E}} \href{http://dx.doi.org/10.1103/PhysRevE.90.052105}{{\bf \bibinfo{volume}{90}}, \bibinfo{pages}{052105}}  (\href{http://dx.doi.org/10.1103/PhysRevE.90.052105}{\bibinfo{year}{2014}}).

\bibitem{khodja_steinigeweg_15}
\bibinfo{author}{A.~Khodja}, \bibinfo{author}{R.~Steinigeweg}, and \bibinfo{author}{J.~Gemmer}, \bibinfo{title}{Relevance of the eigenstate thermalization hypothesis for thermal relaxation}, \bibinfo{journal}{\href{http://dx.doi.org/10.1103/PhysRevE.91.012120}{Phys. Rev. E}} \href{http://dx.doi.org/10.1103/PhysRevE.91.012120}{{\bf \bibinfo{volume}{91}}, \bibinfo{pages}{012120}}  (\href{http://dx.doi.org/10.1103/PhysRevE.91.012120}{\bibinfo{year}{2015}}).

\bibitem{beugeling_moessner_15}
\bibinfo{author}{W.~Beugeling}, \bibinfo{author}{R.~Moessner}, and \bibinfo{author}{M.~Haque}, \bibinfo{title}{Off-diagonal matrix elements of local operators in many-body quantum systems}, \bibinfo{journal}{\href{http://dx.doi.org/10.1103/PhysRevE.91.012144}{Phys. Rev. E}} \href{http://dx.doi.org/10.1103/PhysRevE.91.012144}{{\bf \bibinfo{volume}{91}}, \bibinfo{pages}{012144}}  (\href{http://dx.doi.org/10.1103/PhysRevE.91.012144}{\bibinfo{year}{2015}}).

\bibitem{mondaini_fratus_16}
\bibinfo{author}{R.~Mondaini}, \bibinfo{author}{K.~R. Fratus}, \bibinfo{author}{M.~Srednicki}, and \bibinfo{author}{M.~Rigol}, \bibinfo{title}{{Eigenstate thermalization in the two-dimensional transverse field Ising model}}, \bibinfo{journal}{\href{http://dx.doi.org/10.1103/PhysRevE.93.032104}{Phys. Rev. E}} \href{http://dx.doi.org/10.1103/PhysRevE.93.032104}{{\bf \bibinfo{volume}{93}}, \bibinfo{pages}{032104}}  (\href{http://dx.doi.org/10.1103/PhysRevE.93.032104}{\bibinfo{year}{2016}}).

\bibitem{lan_powell_17}
\bibinfo{author}{Z.~Lan} and \bibinfo{author}{S.~Powell}, \bibinfo{title}{Eigenstate thermalization hypothesis in quantum dimer models}, \bibinfo{journal}{\href{http://dx.doi.org/10.1103/PhysRevB.96.115140}{Phys. Rev. B}} \href{http://dx.doi.org/10.1103/PhysRevB.96.115140}{{\bf \bibinfo{volume}{96}}, \bibinfo{pages}{115140}}  (\href{http://dx.doi.org/10.1103/PhysRevB.96.115140}{\bibinfo{year}{2017}}).

\bibitem{mondaini_rigol_17}
\bibinfo{author}{R.~Mondaini} and \bibinfo{author}{M.~Rigol}, \bibinfo{title}{{Eigenstate thermalization in the two-dimensional transverse field Ising model. II. Off-diagonal matrix elements of observables}}, \bibinfo{journal}{\href{http://dx.doi.org/10.1103/PhysRevE.96.012157}{Phys. Rev. E}} \href{http://dx.doi.org/10.1103/PhysRevE.96.012157}{{\bf \bibinfo{volume}{96}}, \bibinfo{pages}{012157}}  (\href{http://dx.doi.org/10.1103/PhysRevE.96.012157}{\bibinfo{year}{2017}}).

\bibitem{nation_porras_18}
\bibinfo{author}{C.~Nation} and \bibinfo{author}{D.~Porras}, \bibinfo{title}{Off-diagonal observable elements from random matrix theory: distributions, fluctuations, and eigenstate thermalization}, \bibinfo{journal}{\href{http://dx.doi.org/10.1088/1367-2630/aae28f}{New J. Phys.}} \href{http://dx.doi.org/10.1088/1367-2630/aae28f}{{\bf \bibinfo{volume}{20}}, \bibinfo{pages}{103003}}  (\href{http://dx.doi.org/10.1088/1367-2630/aae28f}{\bibinfo{year}{2018}}).

\bibitem{yoshizawa_iyoda_18}
\bibinfo{author}{T.~Yoshizawa}, \bibinfo{author}{E.~Iyoda}, and \bibinfo{author}{T.~Sagawa}, \bibinfo{title}{{Numerical Large Deviation Analysis of the Eigenstate Thermalization Hypothesis}}, \bibinfo{journal}{\href{http://dx.doi.org/10.1103/PhysRevLett.120.200604}{Phys. Rev. Lett.}} \href{http://dx.doi.org/10.1103/PhysRevLett.120.200604}{{\bf \bibinfo{volume}{120}}, \bibinfo{pages}{200604}}  (\href{http://dx.doi.org/10.1103/PhysRevLett.120.200604}{\bibinfo{year}{2018}}).

\bibitem{jansen_stolpp_19}
\bibinfo{author}{D.~Jansen}, \bibinfo{author}{J.~Stolpp}, \bibinfo{author}{L.~Vidmar}, and \bibinfo{author}{F.~Heidrich-Meisner}, \bibinfo{title}{{Eigenstate thermalization and quantum chaos in the Holstein polaron model}}, \bibinfo{journal}{\href{http://dx.doi.org/10.1103/PhysRevB.99.155130}{Phys. Rev. B}} \href{http://dx.doi.org/10.1103/PhysRevB.99.155130}{{\bf \bibinfo{volume}{99}}, \bibinfo{pages}{155130}}  (\href{http://dx.doi.org/10.1103/PhysRevB.99.155130}{\bibinfo{year}{2019}}).

\bibitem{khaymovich_haque_19}
\bibinfo{author}{I.~M. Khaymovich}, \bibinfo{author}{M.~Haque}, and \bibinfo{author}{P.~A. McClarty}, \bibinfo{title}{{Eigenstate Thermalization, Random Matrix Theory, and Behemoths}}, \bibinfo{journal}{\href{http://dx.doi.org/10.1103/PhysRevLett.122.070601}{Phys. Rev. Lett.}} \href{http://dx.doi.org/10.1103/PhysRevLett.122.070601}{{\bf \bibinfo{volume}{122}}, \bibinfo{pages}{070601}}  (\href{http://dx.doi.org/10.1103/PhysRevLett.122.070601}{\bibinfo{year}{2019}}).

\bibitem{leblond_mallayya_19}
\bibinfo{author}{T.~LeBlond}, \bibinfo{author}{K.~Mallayya}, \bibinfo{author}{L.~Vidmar}, and \bibinfo{author}{M.~Rigol}, \bibinfo{title}{Entanglement and matrix elements of observables in interacting integrable systems}, \bibinfo{journal}{\href{http://dx.doi.org/10.1103/PhysRevE.100.062134}{Phys. Rev. E}} \href{http://dx.doi.org/10.1103/PhysRevE.100.062134}{{\bf \bibinfo{volume}{100}}, \bibinfo{pages}{062134}}  (\href{http://dx.doi.org/10.1103/PhysRevE.100.062134}{\bibinfo{year}{2019}}).

\bibitem{Mierzejewski_2020}
\bibinfo{author}{M.~Mierzejewski} and \bibinfo{author}{L.~Vidmar}, \bibinfo{title}{Quantitative impact of integrals of motion on the eigenstate thermalization hypothesis}, \bibinfo{journal}{\href{http://dx.doi.org/10.1103/PhysRevLett.124.040603}{Phys. Rev. Lett.}} \href{http://dx.doi.org/10.1103/PhysRevLett.124.040603}{{\bf \bibinfo{volume}{124}}, \bibinfo{pages}{040603}}  (\href{http://dx.doi.org/10.1103/PhysRevLett.124.040603}{\bibinfo{year}{2020}}).

\bibitem{brenes_leblond_20}
\bibinfo{author}{M.~Brenes}, \bibinfo{author}{T.~LeBlond}, \bibinfo{author}{J.~Goold}, and \bibinfo{author}{M.~Rigol}, \bibinfo{title}{{Eigenstate Thermalization in a Locally Perturbed Integrable System}}, \bibinfo{journal}{\href{http://dx.doi.org/10.1103/PhysRevLett.125.070605}{Phys. Rev. Lett.}} \href{http://dx.doi.org/10.1103/PhysRevLett.125.070605}{{\bf \bibinfo{volume}{125}}, \bibinfo{pages}{070605}}  (\href{http://dx.doi.org/10.1103/PhysRevLett.125.070605}{\bibinfo{year}{2020}}).

\bibitem{brenes_goold_20}
\bibinfo{author}{M.~Brenes}, \bibinfo{author}{J.~Goold}, and \bibinfo{author}{M.~Rigol}, \bibinfo{title}{{Low-frequency behavior of off-diagonal matrix elements in the integrable XXZ chain and in a locally perturbed quantum-chaotic XXZ chain}}, \bibinfo{journal}{\href{http://dx.doi.org/10.1103/PhysRevB.102.075127}{Phys. Rev. B}} \href{http://dx.doi.org/10.1103/PhysRevB.102.075127}{{\bf \bibinfo{volume}{102}}, \bibinfo{pages}{075127}}  (\href{http://dx.doi.org/10.1103/PhysRevB.102.075127}{\bibinfo{year}{2020}}).

\bibitem{leblond_rigol_20}
\bibinfo{author}{T.~LeBlond} and \bibinfo{author}{M.~Rigol}, \bibinfo{title}{{Eigenstate thermalization for observables that break Hamiltonian symmetries and its counterpart in interacting integrable systems}}, \bibinfo{journal}{\href{http://dx.doi.org/10.1103/PhysRevE.102.062113}{Phys. Rev. E}} \href{http://dx.doi.org/10.1103/PhysRevE.102.062113}{{\bf \bibinfo{volume}{102}}, \bibinfo{pages}{062113}}  (\href{http://dx.doi.org/10.1103/PhysRevE.102.062113}{\bibinfo{year}{2020}}).

\bibitem{richter_dymarsky_20}
\bibinfo{author}{J.~Richter}, \bibinfo{author}{A.~Dymarsky}, \bibinfo{author}{R.~Steinigeweg}, and \bibinfo{author}{J.~Gemmer}, \bibinfo{title}{Eigenstate thermalization hypothesis beyond standard indicators: Emergence of random-matrix behavior at small frequencies}, \bibinfo{journal}{\href{http://dx.doi.org/10.1103/PhysRevE.102.042127}{Phys. Rev. E}} \href{http://dx.doi.org/10.1103/PhysRevE.102.042127}{{\bf \bibinfo{volume}{102}}, \bibinfo{pages}{042127}}  (\href{http://dx.doi.org/10.1103/PhysRevE.102.042127}{\bibinfo{year}{2020}}).

\bibitem{wang_lamann_22}
\bibinfo{author}{J.~Wang}, \bibinfo{author}{M.~H. Lamann}, \bibinfo{author}{J.~Richter}, \bibinfo{author}{R.~Steinigeweg}, \bibinfo{author}{A.~Dymarsky}, and \bibinfo{author}{J.~Gemmer}, \bibinfo{title}{Eigenstate thermalization hypothesis and its deviations from random-matrix theory beyond the thermalization time}, \bibinfo{journal}{\href{http://dx.doi.org/10.1103/PhysRevLett.128.180601}{Phys. Rev. Lett.}} \href{http://dx.doi.org/10.1103/PhysRevLett.128.180601}{{\bf \bibinfo{volume}{128}}, \bibinfo{pages}{180601}}  (\href{http://dx.doi.org/10.1103/PhysRevLett.128.180601}{\bibinfo{year}{2022}}).

\bibitem{noh_21}
\bibinfo{author}{J.~D. Noh}, \bibinfo{title}{Eigenstate thermalization hypothesis and eigenstate-to-eigenstate fluctuations}, \bibinfo{journal}{\href{http://dx.doi.org/10.1103/PhysRevE.103.012129}{Phys. Rev. E}} \href{http://dx.doi.org/10.1103/PhysRevE.103.012129}{{\bf \bibinfo{volume}{103}}, \bibinfo{pages}{012129}}  (\href{http://dx.doi.org/10.1103/PhysRevE.103.012129}{\bibinfo{year}{2021}}).

\bibitem{schoenle_jansen_21}
\bibinfo{author}{C.~Sch\"onle}, \bibinfo{author}{D.~Jansen}, \bibinfo{author}{F.~Heidrich-Meisner}, and \bibinfo{author}{L.~Vidmar}, \bibinfo{title}{Eigenstate thermalization hypothesis through the lens of autocorrelation functions}, \bibinfo{journal}{\href{http://dx.doi.org/10.1103/PhysRevB.103.235137}{Phys. Rev. B}} \href{http://dx.doi.org/10.1103/PhysRevB.103.235137}{{\bf \bibinfo{volume}{103}}, \bibinfo{pages}{235137}}  (\href{http://dx.doi.org/10.1103/PhysRevB.103.235137}{\bibinfo{year}{2021}}).

\bibitem{sugimoto_hamazaki_21}
\bibinfo{author}{S.~Sugimoto}, \bibinfo{author}{R.~Hamazaki}, and \bibinfo{author}{M.~Ueda}, \bibinfo{title}{Test of the eigenstate thermalization hypothesis based on local random matrix theory}, \bibinfo{journal}{\href{http://dx.doi.org/10.1103/PhysRevLett.126.120602}{Phys. Rev. Lett.}} \href{http://dx.doi.org/10.1103/PhysRevLett.126.120602}{{\bf \bibinfo{volume}{126}}, \bibinfo{pages}{120602}}  (\href{http://dx.doi.org/10.1103/PhysRevLett.126.120602}{\bibinfo{year}{2021}}).

\bibitem{noh_23}
\bibinfo{author}{J.~D. Noh}, \bibinfo{title}{Eigenstate thermalization hypothesis in two-dimensional $xxz$ model with or without su(2) symmetry}, \bibinfo{journal}{\href{http://dx.doi.org/10.1103/PhysRevE.107.014130}{Phys. Rev. E}} \href{http://dx.doi.org/10.1103/PhysRevE.107.014130}{{\bf \bibinfo{volume}{107}}, \bibinfo{pages}{014130}}  (\href{http://dx.doi.org/10.1103/PhysRevE.107.014130}{\bibinfo{year}{2023}}).

\bibitem{wang_zhu_24}
\bibinfo{author}{D.-Z. Wang}, \bibinfo{author}{H.~Zhu}, \bibinfo{author}{J.~Cui}, \bibinfo{author}{J.~Arg\"uello-Luengo}, \bibinfo{author}{M.~Lewenstein}, \bibinfo{author}{G.-F. Zhang}, \bibinfo{author}{P.~Sierant}, and \bibinfo{author}{S.-J. Ran}, \bibinfo{title}{Eigenstate thermalization and its breakdown in quantum spin chains with inhomogeneous interactions}, \bibinfo{journal}{\href{http://dx.doi.org/10.1103/PhysRevB.109.045139}{Phys. Rev. B}} \href{http://dx.doi.org/10.1103/PhysRevB.109.045139}{{\bf \bibinfo{volume}{109}}, \bibinfo{pages}{045139}}  (\href{http://dx.doi.org/10.1103/PhysRevB.109.045139}{\bibinfo{year}{2024}}).

\bibitem{luo_trivedi_24}
\bibinfo{author}{M.~Luo}, \bibinfo{author}{R.~Trivedi}, \bibinfo{author}{M.~C. Ba\~nuls}, and \bibinfo{author}{J.~I. Cirac}, \bibinfo{title}{Probing off-diagonal eigenstate thermalization with tensor networks}, \bibinfo{journal}{\href{http://dx.doi.org/10.1103/PhysRevB.109.134304}{Phys. Rev. B}} \href{http://dx.doi.org/10.1103/PhysRevB.109.134304}{{\bf \bibinfo{volume}{109}}, \bibinfo{pages}{134304}}  (\href{http://dx.doi.org/10.1103/PhysRevB.109.134304}{\bibinfo{year}{2024}}).

\bibitem{ebner_schafer_24}
\bibinfo{author}{L.~Ebner}, \bibinfo{author}{A.~Sch\"afer}, \bibinfo{author}{C.~Seidl}, \bibinfo{author}{B.~M\"uller}, and \bibinfo{author}{X.~Yao}, \bibinfo{title}{Eigenstate thermalization in ($2+1$)-dimensional su(2) lattice gauge theory}, \bibinfo{journal}{\href{http://dx.doi.org/10.1103/PhysRevD.109.014504}{Phys. Rev. D}} \href{http://dx.doi.org/10.1103/PhysRevD.109.014504}{{\bf \bibinfo{volume}{109}}, \bibinfo{pages}{014504}}  (\href{http://dx.doi.org/10.1103/PhysRevD.109.014504}{\bibinfo{year}{2024}}).

\bibitem{patil_rigol_25}
\bibinfo{author}{R.~Patil} and \bibinfo{author}{M.~Rigol}, \bibinfo{title}{Eigenstate thermalization in spin-$\frac{1}{2}$ systems with su(2) symmetry}, \bibinfo{journal}{\href{http://dx.doi.org/10.1103/PhysRevB.111.205126}{Phys. Rev. B}} \href{http://dx.doi.org/10.1103/PhysRevB.111.205126}{{\bf \bibinfo{volume}{111}}, \bibinfo{pages}{205126}}  (\href{http://dx.doi.org/10.1103/PhysRevB.111.205126}{\bibinfo{year}{2025}}).

\bibitem{capizzi_wang_25}
\bibinfo{author}{L.~Capizzi}, \bibinfo{author}{J.~Wang}, \bibinfo{author}{X.~Xu}, \bibinfo{author}{L.~Mazza}, and \bibinfo{author}{D.~Poletti}, \bibinfo{title}{Hydrodynamics and the eigenstate thermalization hypothesis}, \bibinfo{journal}{\href{http://dx.doi.org/10.1103/PhysRevX.15.011059}{Phys. Rev. X}} \href{http://dx.doi.org/10.1103/PhysRevX.15.011059}{{\bf \bibinfo{volume}{15}}, \bibinfo{pages}{011059}}  (\href{http://dx.doi.org/10.1103/PhysRevX.15.011059}{\bibinfo{year}{2025}}).

\bibitem{saiaramthottil_emamikopaei_25}
\bibinfo{author}{A.~Sai~Aramthottil}, \bibinfo{author}{A.~Emami~Kopaei}, \bibinfo{author}{P.~Sierant}, \bibinfo{author}{L.~Vidmar}, and \bibinfo{author}{J.~Zakrzewski}, \bibinfo{title}{False signatures of non-ergodic behavior in disordered quantum many-body systems}, \bibinfo{journal}{\href{http://dx.doi.org/10.1088/1367-2630/ae20b3}{New J. Phys.}} \href{http://dx.doi.org/10.1088/1367-2630/ae20b3}{{\bf \bibinfo{volume}{27}}, \bibinfo{pages}{125001}}  (\href{http://dx.doi.org/10.1088/1367-2630/ae20b3}{\bibinfo{year}{2025}}).

\bibitem{srednicki_99}
\bibinfo{author}{M.~Srednicki}, \bibinfo{title}{The approach to thermal equilibrium in quantized chaotic systems}, \bibinfo{journal}{\href{http://dx.doi.org/10.1088/0305-4470/32/7/007}{J. Phys. A.}} \href{http://dx.doi.org/10.1088/0305-4470/32/7/007}{{\bf \bibinfo{volume}{32}}, \bibinfo{pages}{1163}}  (\href{http://dx.doi.org/10.1088/0305-4470/32/7/007}{\bibinfo{year}{1999}}).

\bibitem{murthy_srednicki_19b}
\bibinfo{author}{C.~Murthy} and \bibinfo{author}{M.~Srednicki}, \bibinfo{title}{{Bounds on Chaos from the Eigenstate Thermalization Hypothesis}}, \bibinfo{journal}{\href{http://dx.doi.org/10.1103/PhysRevLett.123.230606}{Phys. Rev. Lett.}} \href{http://dx.doi.org/10.1103/PhysRevLett.123.230606}{{\bf \bibinfo{volume}{123}}, \bibinfo{pages}{230606}}  (\href{http://dx.doi.org/10.1103/PhysRevLett.123.230606}{\bibinfo{year}{2019}}).

\bibitem{brenes_pappalardi_21}
\bibinfo{author}{M.~Brenes}, \bibinfo{author}{S.~Pappalardi}, \bibinfo{author}{M.~T. Mitchison}, \bibinfo{author}{J.~Goold}, and \bibinfo{author}{A.~Silva}, \bibinfo{title}{Out-of-time-order correlations and the fine structure of eigenstate thermalization}, \bibinfo{journal}{\href{http://dx.doi.org/10.1103/PhysRevE.104.034120}{Phys. Rev. E}} \href{http://dx.doi.org/10.1103/PhysRevE.104.034120}{{\bf \bibinfo{volume}{104}}, \bibinfo{pages}{034120}}  (\href{http://dx.doi.org/10.1103/PhysRevE.104.034120}{\bibinfo{year}{2021}}).

\bibitem{pappalardi_foini_22}
\bibinfo{author}{S.~Pappalardi}, \bibinfo{author}{L.~Foini}, and \bibinfo{author}{J.~Kurchan}, \bibinfo{title}{Eigenstate thermalization hypothesis and free probability}, \bibinfo{journal}{\href{http://dx.doi.org/10.1103/PhysRevLett.129.170603}{Phys. Rev. Lett.}} \href{http://dx.doi.org/10.1103/PhysRevLett.129.170603}{{\bf \bibinfo{volume}{129}}, \bibinfo{pages}{170603}}  (\href{http://dx.doi.org/10.1103/PhysRevLett.129.170603}{\bibinfo{year}{2022}}).

\bibitem{hahn_luitz_24}
\bibinfo{author}{D.~Hahn}, \bibinfo{author}{D.~J. Luitz}, and \bibinfo{author}{J.~T. Chalker}, \bibinfo{title}{Eigenstate correlations, the eigenstate thermalization hypothesis, and quantum information dynamics in chaotic many-body quantum systems}, \bibinfo{journal}{\href{http://dx.doi.org/10.1103/PhysRevX.14.031029}{Phys. Rev. X}} \href{http://dx.doi.org/10.1103/PhysRevX.14.031029}{{\bf \bibinfo{volume}{14}}, \bibinfo{pages}{031029}}  (\href{http://dx.doi.org/10.1103/PhysRevX.14.031029}{\bibinfo{year}{2024}}).

\bibitem{fava_kurchan_25}
\bibinfo{author}{M.~Fava}, \bibinfo{author}{J.~Kurchan}, and \bibinfo{author}{S.~Pappalardi}, \bibinfo{title}{Designs via free probability}, \bibinfo{journal}{\href{http://dx.doi.org/10.1103/PhysRevX.15.011031}{Phys. Rev. X}} \href{http://dx.doi.org/10.1103/PhysRevX.15.011031}{{\bf \bibinfo{volume}{15}}, \bibinfo{pages}{011031}}  (\href{http://dx.doi.org/10.1103/PhysRevX.15.011031}{\bibinfo{year}{2025}}).

\bibitem{pappalardi_fritzsch_25}
\bibinfo{author}{S.~Pappalardi}, \bibinfo{author}{F.~Fritzsch}, and \bibinfo{author}{T.~Prosen}, \bibinfo{title}{Full eigenstate thermalization via free cumulants in quantum lattice systems}, \bibinfo{journal}{\href{http://dx.doi.org/10.1103/PhysRevLett.134.140404}{Phys. Rev. Lett.}} \href{http://dx.doi.org/10.1103/PhysRevLett.134.140404}{{\bf \bibinfo{volume}{134}}, \bibinfo{pages}{140404}}  (\href{http://dx.doi.org/10.1103/PhysRevLett.134.140404}{\bibinfo{year}{2025}}).

\bibitem{foini_kurchan_19}
\bibinfo{author}{L.~Foini} and \bibinfo{author}{J.~Kurchan}, \bibinfo{title}{Eigenstate thermalization hypothesis and out of time order correlators}, \bibinfo{journal}{\href{http://dx.doi.org/10.1103/PhysRevE.99.042139}{Phys. Rev. E}} \href{http://dx.doi.org/10.1103/PhysRevE.99.042139}{{\bf \bibinfo{volume}{99}}, \bibinfo{pages}{042139}}  (\href{http://dx.doi.org/10.1103/PhysRevE.99.042139}{\bibinfo{year}{2019}}).

\bibitem{chan_deluca_19}
\bibinfo{author}{A.~Chan}, \bibinfo{author}{A.~De~Luca}, and \bibinfo{author}{J.~T. Chalker}, \bibinfo{title}{{Eigenstate Correlations, Thermalization, and the Butterfly Effect}}, \bibinfo{journal}{\href{http://dx.doi.org/10.1103/PhysRevLett.122.220601}{Phys. Rev. Lett.}} \href{http://dx.doi.org/10.1103/PhysRevLett.122.220601}{{\bf \bibinfo{volume}{122}}, \bibinfo{pages}{220601}}  (\href{http://dx.doi.org/10.1103/PhysRevLett.122.220601}{\bibinfo{year}{2019}}).

\bibitem{biroli_kollath_10}
\bibinfo{author}{G.~Biroli}, \bibinfo{author}{C.~Kollath}, and \bibinfo{author}{A.~M. L\"auchli}, \bibinfo{title}{{Effect of Rare Fluctuations on the Thermalization of Isolated Quantum Systems}}, \bibinfo{journal}{\href{http://dx.doi.org/10.1103/PhysRevLett.105.250401}{Phys. Rev. Lett.}} \href{http://dx.doi.org/10.1103/PhysRevLett.105.250401}{{\bf \bibinfo{volume}{105}}, \bibinfo{pages}{250401}}  (\href{http://dx.doi.org/10.1103/PhysRevLett.105.250401}{\bibinfo{year}{2010}}).

\bibitem{Cassidy_2011}
\bibinfo{author}{A.~C. Cassidy}, \bibinfo{author}{C.~W. Clark}, and \bibinfo{author}{M.~Rigol}, \bibinfo{title}{Generalized thermalization in an integrable lattice system}, \bibinfo{journal}{\href{http://dx.doi.org/10.1103/physrevlett.106.140405}{Phys. Rev. Lett.}} \href{http://dx.doi.org/10.1103/physrevlett.106.140405}{{\bf \bibinfo{volume}{106}}, \bibinfo{pages}{140405}}  (\href{http://dx.doi.org/10.1103/physrevlett.106.140405}{\bibinfo{year}{2011}}).

\bibitem{vidmar16}
\bibinfo{author}{L.~Vidmar} and \bibinfo{author}{M.~Rigol}, \bibinfo{title}{{Generalized Gibbs ensemble in integrable lattice models}}, \bibinfo{journal}{\href{http://dx.doi.org/10.1088/1742-5468/2016/06/064007}{J. Stat. Mech.}} \href{http://dx.doi.org/10.1088/1742-5468/2016/06/064007}{{\bf \bibinfo{volume}{{\rm (2016)}}}, \bibinfo{pages}{064007}}.

\bibitem{zhang_vidmar_22}
\bibinfo{author}{Y.~Zhang}, \bibinfo{author}{L.~Vidmar}, and \bibinfo{author}{M.~Rigol}, \bibinfo{title}{Statistical properties of the off-diagonal matrix elements of observables in eigenstates of integrable systems}, \bibinfo{journal}{\href{http://dx.doi.org/10.1103/PhysRevE.106.014132}{Phys. Rev. E}} \href{http://dx.doi.org/10.1103/PhysRevE.106.014132}{{\bf \bibinfo{volume}{106}}, \bibinfo{pages}{014132}}  (\href{http://dx.doi.org/10.1103/PhysRevE.106.014132}{\bibinfo{year}{2022}}).

\bibitem{lydzba_swietek_24}
\bibinfo{author}{P.~\L{}yd\ifmmode~\dot{z}\else \.{z}\fi{}ba}, \bibinfo{author}{R.~\ifmmode \acute{S}\else \'{S}\fi{}wi\ifmmode~\mbox{\k{e}}\else \k{e}\fi{}tek}, \bibinfo{author}{M.~Mierzejewski}, \bibinfo{author}{M.~Rigol}, and \bibinfo{author}{L.~Vidmar}, \bibinfo{title}{Normal weak eigenstate thermalization}, \bibinfo{journal}{\href{http://dx.doi.org/10.1103/PhysRevB.110.104202}{Phys. Rev. B}} \href{http://dx.doi.org/10.1103/PhysRevB.110.104202}{{\bf \bibinfo{volume}{110}}, \bibinfo{pages}{104202}}  (\href{http://dx.doi.org/10.1103/PhysRevB.110.104202}{\bibinfo{year}{2024}}).

\bibitem{Ikeda2013}
\bibinfo{author}{T.~N. Ikeda}, \bibinfo{author}{Y.~Watanabe}, and \bibinfo{author}{M.~Ueda}, \bibinfo{title}{{Finite-size scaling analysis of the eigenstate thermalization hypothesis in a one-dimensional interacting Bose gas}}, \bibinfo{journal}{\href{http://dx.doi.org/10.1103/PhysRevE.87.012125}{Phys. Rev. E}} \href{http://dx.doi.org/10.1103/PhysRevE.87.012125}{{\bf \bibinfo{volume}{87}}, \bibinfo{pages}{012125}}  (\href{http://dx.doi.org/10.1103/PhysRevE.87.012125}{\bibinfo{year}{2013}}).

\bibitem{alba_15}
\bibinfo{author}{V.~Alba}, \bibinfo{title}{Eigenstate thermalization hypothesis and integrability in quantum spin chains}, \bibinfo{journal}{\href{http://dx.doi.org/10.1103/PhysRevB.91.155123}{Phys. Rev. B}} \href{http://dx.doi.org/10.1103/PhysRevB.91.155123}{{\bf \bibinfo{volume}{91}}, \bibinfo{pages}{155123}}  (\href{http://dx.doi.org/10.1103/PhysRevB.91.155123}{\bibinfo{year}{2015}}).

\bibitem{mori_16}
\bibinfo{author}{T.~{Mori}}, \bibinfo{title}{{Weak eigenstate thermalization with large deviation bound}}, \bibinfo{journal}{\href{http://dx.doi.org/10.48550/arXiv.1609.09776}{arXiv e-prints}} \href{http://dx.doi.org/10.48550/arXiv.1609.09776}{\bibinfo{eid}{arXiv:1609.09776}}  (\href{http://dx.doi.org/10.48550/arXiv.1609.09776}{\bibinfo{year}{2016}}). \eprint{1609.09776}.

\bibitem{essler_deklerk_24}
\bibinfo{author}{F.~H.~L. Essler} and \bibinfo{author}{A.~J. J.~M. de~Klerk}, \bibinfo{title}{Statistics of matrix elements of local operators in integrable models}, \bibinfo{journal}{\href{http://dx.doi.org/10.1103/PhysRevX.14.031048}{Phys. Rev. X}} \href{http://dx.doi.org/10.1103/PhysRevX.14.031048}{{\bf \bibinfo{volume}{14}}, \bibinfo{pages}{031048}}  (\href{http://dx.doi.org/10.1103/PhysRevX.14.031048}{\bibinfo{year}{2024}}).

\bibitem{shiraishi_mori_17}
\bibinfo{author}{N.~Shiraishi} and \bibinfo{author}{T.~Mori}, \bibinfo{title}{Systematic construction of counterexamples to the eigenstate thermalization hypothesis}, \bibinfo{journal}{\href{http://dx.doi.org/10.1103/PhysRevLett.119.030601}{Phys. Rev. Lett.}} \href{http://dx.doi.org/10.1103/PhysRevLett.119.030601}{{\bf \bibinfo{volume}{119}}, \bibinfo{pages}{030601}}  (\href{http://dx.doi.org/10.1103/PhysRevLett.119.030601}{\bibinfo{year}{2017}}).

\bibitem{turner_michailidis_18b}
\bibinfo{author}{C.~J. Turner}, \bibinfo{author}{A.~A. Michailidis}, \bibinfo{author}{D.~A. Abanin}, \bibinfo{author}{M.~Serbyn}, and \bibinfo{author}{Z.~Papi\ifmmode~\acute{c}\else \'{c}\fi{}}, \bibinfo{title}{{Quantum scarred eigenstates in a Rydberg atom chain: Entanglement, breakdown of thermalization, and stability to perturbations}}, \bibinfo{journal}{\href{http://dx.doi.org/10.1103/PhysRevB.98.155134}{Phys. Rev. B}} \href{http://dx.doi.org/10.1103/PhysRevB.98.155134}{{\bf \bibinfo{volume}{98}}, \bibinfo{pages}{155134}}  (\href{http://dx.doi.org/10.1103/PhysRevB.98.155134}{\bibinfo{year}{2018}}).

\bibitem{dooley_kells_22}
\bibinfo{author}{S.~Dooley} and \bibinfo{author}{G.~Kells}, \bibinfo{title}{Extreme many-body scarring in a quantum spin chain via weak dynamical constraints}, \bibinfo{journal}{\href{http://dx.doi.org/10.1103/PhysRevB.105.155127}{Phys. Rev. B}} \href{http://dx.doi.org/10.1103/PhysRevB.105.155127}{{\bf \bibinfo{volume}{105}}, \bibinfo{pages}{155127}}  (\href{http://dx.doi.org/10.1103/PhysRevB.105.155127}{\bibinfo{year}{2022}}).

\bibitem{wang_zhou_25}
\bibinfo{author}{J.-w. Wang}, \bibinfo{author}{X.-F. Zhou}, \bibinfo{author}{G.-C. Guo}, and \bibinfo{author}{Z.-W. Zhou}, \bibinfo{title}{Thermalization of quantum many-body scars in kinetically constrained systems}, \bibinfo{journal}{\href{http://dx.doi.org/10.1103/sf1z-yqyk}{Phys. Rev. Res.}} \href{http://dx.doi.org/10.1103/sf1z-yqyk}{{\bf \bibinfo{volume}{7}}, \bibinfo{pages}{043337}}  (\href{http://dx.doi.org/10.1103/sf1z-yqyk}{\bibinfo{year}{2025}}).

\bibitem{moudgalya_prem_21}
\bibinfo{author}{S.~Moudgalya}, \bibinfo{author}{A.~Prem}, \bibinfo{author}{R.~Nandkishore}, \bibinfo{author}{N.~Regnault}, and \bibinfo{author}{B.~A. Bernevig}, \bibinfo{title}{Thermalization and its absence within krylov subspaces of a constrained hamiltonian}, \bibinfo{howpublished}{in \textit{Memorial Volume for Shoucheng Zhang} (World Scientific, Singapore, 2021), Chap. 7, pp. 147--209}.

\bibitem{aditya_dhar_24}
\bibinfo{author}{S.~Aditya}, \bibinfo{author}{D.~Dhar}, and \bibinfo{author}{D.~Sen}, \bibinfo{title}{Subspace-restricted thermalization in a correlated-hopping model with strong hilbert space fragmentation characterized by irreducible strings}, \bibinfo{journal}{\href{http://dx.doi.org/10.1103/PhysRevB.110.045418}{Phys. Rev. B}} \href{http://dx.doi.org/10.1103/PhysRevB.110.045418}{{\bf \bibinfo{volume}{110}}, \bibinfo{pages}{045418}}  (\href{http://dx.doi.org/10.1103/PhysRevB.110.045418}{\bibinfo{year}{2024}}).

\bibitem{kliczkowski_swietek_24}
\bibinfo{author}{M.~Kliczkowski}, \bibinfo{author}{R.~\ifmmode \acute{S}\else \'{S}\fi{}wi\ifmmode~\mbox{\k{e}}\else \k{e}\fi{}tek}, \bibinfo{author}{M.~Hopjan}, and \bibinfo{author}{L.~Vidmar}, \bibinfo{title}{Fading ergodicity}, \bibinfo{journal}{\href{http://dx.doi.org/10.1103/PhysRevB.110.134206}{Phys. Rev. B}} \href{http://dx.doi.org/10.1103/PhysRevB.110.134206}{{\bf \bibinfo{volume}{110}}, \bibinfo{pages}{134206}}  (\href{http://dx.doi.org/10.1103/PhysRevB.110.134206}{\bibinfo{year}{2024}}).

\bibitem{swietek_lydzba_25}
\bibinfo{author}{R.~\ifmmode \acute{S}\else \'{S}\fi{}wi\ifmmode~\mbox{\c{e}}\else \c{e}\fi{}tek}, \bibinfo{author}{P.~\L{}yd\ifmmode~\dot{z}\else \.{z}\fi{}ba}, and \bibinfo{author}{L.~Vidmar}, \bibinfo{title}{Fading ergodicity meets maximal chaos}, \bibinfo{journal}{\href{http://dx.doi.org/10.1103/PhysRevB.111.184203}{Phys. Rev. B}} \href{http://dx.doi.org/10.1103/PhysRevB.111.184203}{{\bf \bibinfo{volume}{111}}, \bibinfo{pages}{184203}}  (\href{http://dx.doi.org/10.1103/PhysRevB.111.184203}{\bibinfo{year}{2025}}).

\bibitem{swietek_hopjan_25}
\bibinfo{author}{R.~\ifmmode \acute{S}\else \'{S}\fi{}wi\ifmmode~\mbox{\k{e}}\else \k{e}\fi{}tek}, \bibinfo{author}{M.~Hopjan}, \bibinfo{author}{C.~Vanoni}, \bibinfo{author}{A.~Scardicchio}, and \bibinfo{author}{L.~Vidmar}, \bibinfo{title}{Scaling theory of fading ergodicity}, \bibinfo{journal}{\href{http://dx.doi.org/10.1103/4l42-l7pk}{Phys. Rev. Lett.}} \href{http://dx.doi.org/10.1103/4l42-l7pk}{{\bf \bibinfo{volume}{135}}, \bibinfo{pages}{170401}}  (\href{http://dx.doi.org/10.1103/4l42-l7pk}{\bibinfo{year}{2025}}).

\bibitem{swietek_kliczkowski_26}
\bibinfo{author}{R.~\ifmmode \acute{S}\else \'{S}\fi{}wi\ifmmode~\mbox{\k{e}}\else \k{e}\fi{}tek}, \bibinfo{author}{M.~Kliczkowski}, \bibinfo{author}{M.~Hopjan}, and \bibinfo{author}{L.~Vidmar}, \bibinfo{title}{Fading ergodicity and quantum dynamics in random matrix ensembles}, \bibinfo{journal}{\href{http://dx.doi.org/10.1103/kzrk-z6nt}{Phys. Rev. B}} \href{http://dx.doi.org/10.1103/kzrk-z6nt}{{\bf \bibinfo{volume}{114}}, \bibinfo{pages}{024207}}  (\href{http://dx.doi.org/10.1103/kzrk-z6nt}{\bibinfo{year}{2026}}).

\bibitem{luitz_barlev_16}
\bibinfo{author}{D.~J. Luitz} and \bibinfo{author}{Y.~Bar~Lev}, \bibinfo{title}{Anomalous thermalization in ergodic systems}, \bibinfo{journal}{\href{http://dx.doi.org/10.1103/PhysRevLett.117.170404}{Phys. Rev. Lett.}} \href{http://dx.doi.org/10.1103/PhysRevLett.117.170404}{{\bf \bibinfo{volume}{117}}, \bibinfo{pages}{170404}}  (\href{http://dx.doi.org/10.1103/PhysRevLett.117.170404}{\bibinfo{year}{2016}}).

\bibitem{balachandran_santos_23}
\bibinfo{author}{V.~Balachandran}, \bibinfo{author}{L.~F. Santos}, \bibinfo{author}{M.~Rigol}, and \bibinfo{author}{D.~Poletti}, \bibinfo{title}{Slow relaxation of out-of-time-ordered correlators in interacting integrable and nonintegrable spin-$\frac{1}{2}$ xyz chains}, \bibinfo{journal}{\href{http://dx.doi.org/10.1103/PhysRevB.107.235421}{Phys. Rev. B}} \href{http://dx.doi.org/10.1103/PhysRevB.107.235421}{{\bf \bibinfo{volume}{107}}, \bibinfo{pages}{235421}}  (\href{http://dx.doi.org/10.1103/PhysRevB.107.235421}{\bibinfo{year}{2023}}).

\bibitem{vidmar_21}
\bibinfo{author}{L.~Vidmar}, \bibinfo{author}{B.~Krajewski}, \bibinfo{author}{J.~Bon\ifmmode~\check{c}\else \v{c}\fi{}a}, and \bibinfo{author}{M.~Mierzejewski}, \bibinfo{title}{{Phenomenology of Spectral Functions in Disordered Spin Chains at Infinite Temperature}}, \bibinfo{journal}{\href{http://dx.doi.org/10.1103/PhysRevLett.127.230603}{Phys. Rev. Lett.}} \href{http://dx.doi.org/10.1103/PhysRevLett.127.230603}{{\bf \bibinfo{volume}{127}}, \bibinfo{pages}{230603}}  (\href{http://dx.doi.org/10.1103/PhysRevLett.127.230603}{\bibinfo{year}{2021}}).

\bibitem{sels_polkovnikov_21}
\bibinfo{author}{D.~Sels} and \bibinfo{author}{A.~Polkovnikov}, \bibinfo{title}{Dynamical obstruction to localization in a disordered spin chain}, \bibinfo{journal}{\href{http://dx.doi.org/10.1103/PhysRevE.104.054105}{Phys. Rev. E}} \href{http://dx.doi.org/10.1103/PhysRevE.104.054105}{{\bf \bibinfo{volume}{104}}, \bibinfo{pages}{054105}}  (\href{http://dx.doi.org/10.1103/PhysRevE.104.054105}{\bibinfo{year}{2021}}).

\bibitem{sels_polkovnikov_23}
\bibinfo{author}{D.~Sels} and \bibinfo{author}{A.~Polkovnikov}, \bibinfo{title}{Thermalization of dilute impurities in one-dimensional spin chains}, \bibinfo{journal}{\href{http://dx.doi.org/10.1103/PhysRevX.13.011041}{Phys. Rev. X}} \href{http://dx.doi.org/10.1103/PhysRevX.13.011041}{{\bf \bibinfo{volume}{13}}, \bibinfo{pages}{011041}}  (\href{http://dx.doi.org/10.1103/PhysRevX.13.011041}{\bibinfo{year}{2023}}).

\bibitem{suntajs_22}
\bibinfo{author}{J.~\ifmmode~\check{S}\else \v{S}\fi{}untajs} and \bibinfo{author}{L.~Vidmar}, \bibinfo{title}{Ergodicity breaking transition in zero dimensions}, \bibinfo{journal}{\href{http://dx.doi.org/10.1103/PhysRevLett.129.060602}{Phys. Rev. Lett.}} \href{http://dx.doi.org/10.1103/PhysRevLett.129.060602}{{\bf \bibinfo{volume}{129}}, \bibinfo{pages}{060602}}  (\href{http://dx.doi.org/10.1103/PhysRevLett.129.060602}{\bibinfo{year}{2022}}).

\bibitem{Note1}
\bibinfo{note}{The square of the Hilbert--Schmidt norm is, for structureless observables, defined as $||\protect \hat A||^2 = {\protect \rm Tr}\{\protect \hat A^2\}/{\protect \cal D}$, see also Ref.~\cite {lydzba_swietek_24}. Normalized observables satisfy $||\protect \hat A||=1$.}

\bibitem{Note2}
\bibinfo{note}{In the language of Ref.~\cite {kliczkowski_swietek_24} that introduced the parameter $\eta $ to characterize fading ergodicity, $1-\theta = 2/\eta $.}

\bibitem{suntajs_20_a}
\bibinfo{author}{J.~\ifmmode~\check{S}\else \v{S}\fi{}untajs}, \bibinfo{author}{J.~Bon\ifmmode~\check{c}\else \v{c}\fi{}a}, \bibinfo{author}{T.~Prosen}, and \bibinfo{author}{L.~Vidmar}, \bibinfo{title}{Quantum chaos challenges many-body localization}, \bibinfo{journal}{\href{http://dx.doi.org/10.1103/PhysRevE.102.062144}{Phys. Rev. E}} \href{http://dx.doi.org/10.1103/PhysRevE.102.062144}{{\bf \bibinfo{volume}{102}}, \bibinfo{pages}{062144}}  (\href{http://dx.doi.org/10.1103/PhysRevE.102.062144}{\bibinfo{year}{2020}}).

\bibitem{suntajs_20_b}
\bibinfo{author}{J.~\ifmmode~\check{S}\else \v{S}\fi{}untajs}, \bibinfo{author}{J.~Bon\ifmmode~\check{c}\else \v{c}\fi{}a}, \bibinfo{author}{T.~Prosen}, and \bibinfo{author}{L.~Vidmar}, \bibinfo{title}{Ergodicity breaking transition in finite disordered spin chains}, \bibinfo{journal}{\href{http://dx.doi.org/10.1103/PhysRevB.102.064207}{Phys. Rev. B}} \href{http://dx.doi.org/10.1103/PhysRevB.102.064207}{{\bf \bibinfo{volume}{102}}, \bibinfo{pages}{064207}}  (\href{http://dx.doi.org/10.1103/PhysRevB.102.064207}{\bibinfo{year}{2020}}).

\bibitem{sierant_lewenstein_20}
\bibinfo{author}{P.~Sierant}, \bibinfo{author}{M.~Lewenstein}, and \bibinfo{author}{J.~Zakrzewski}, \bibinfo{title}{Polynomially filtered exact diagonalization approach to many-body localization}, \bibinfo{journal}{\href{http://dx.doi.org/10.1103/PhysRevLett.125.156601}{Phys. Rev. Lett.}} \href{http://dx.doi.org/10.1103/PhysRevLett.125.156601}{{\bf \bibinfo{volume}{125}}, \bibinfo{pages}{156601}}  (\href{http://dx.doi.org/10.1103/PhysRevLett.125.156601}{\bibinfo{year}{2020}}).

\bibitem{pintar_pawlik_26}
\bibinfo{author}{R.~Pintar}, \bibinfo{author}{K.~Pawlik}, \bibinfo{author}{R.~Świętek}, \bibinfo{author}{M.~Hopjan}, \bibinfo{author}{J.~Šuntajs}, \bibinfo{author}{J.~Zakrzewski}, \bibinfo{author}{P.~Sierant}, and \bibinfo{author}{L.~Vidmar}, \bibinfo{title}{Computing eigenpairs of quantum many-body systems with polfed.jl}, \href{https://arxiv.org/abs/2605.10191}{\bibinfo{howpublished}{arXiv:2605.10191}}. \bibinfo{note}{Code available at \url{https://github.com/RockClimbingRocks/Polfed.jl}}.

\bibitem{mace_19}
\bibinfo{author}{N.~Mac\'e}, \bibinfo{author}{F.~Alet}, and \bibinfo{author}{N.~Laflorencie}, \bibinfo{title}{Multifractal scalings across the many-body localization transition}, \bibinfo{journal}{\href{http://dx.doi.org/10.1103/PhysRevLett.123.180601}{Phys. Rev. Lett.}} \href{http://dx.doi.org/10.1103/PhysRevLett.123.180601}{{\bf \bibinfo{volume}{123}}, \bibinfo{pages}{180601}}  (\href{http://dx.doi.org/10.1103/PhysRevLett.123.180601}{\bibinfo{year}{2019}}).

\bibitem{vega1}
\bibinfo{note}{{www.hpc-rivr.si}}.

\bibitem{vega2}
\bibinfo{note}{{eurohpc-ju.europa.eu}}.

\bibitem{vega3}
\bibinfo{note}{{ www.izum.si}}.

\bibitem{ohgane_2026_23983056}
\bibinfo{author}{K.~Ohgane} and \bibinfo{author}{L.~Vidmar}, \href{https://doi.org/10.5281/zenodo.23983056}{\bibinfo{title}{Data for "two-parameter ansatz for the violation of eigenstate thermalization"}}  (\href{http://dx.doi.org/10.5281/zenodo.23983056}{\bibinfo{year}{2026}}).

\bibitem{nieuwenburg_baum_19}
\bibinfo{author}{E.~van Nieuwenburg}, \bibinfo{author}{Y.~Baum}, and \bibinfo{author}{G.~Refael}, \bibinfo{title}{From bloch oscillations to many-body localization in clean interacting systems}, \bibinfo{journal}{\href{http://dx.doi.org/10.1073/pnas.1819316116}{Proc. Natl. Acad. Sci. U.S.A.}} \href{http://dx.doi.org/10.1073/pnas.1819316116}{{\bf \bibinfo{volume}{116}}, \bibinfo{pages}{9269}}  (\href{http://dx.doi.org/10.1073/pnas.1819316116}{\bibinfo{year}{2019}}).

\bibitem{schulz_hooley_19}
\bibinfo{author}{M.~Schulz}, \bibinfo{author}{C.~A. Hooley}, \bibinfo{author}{R.~Moessner}, and \bibinfo{author}{F.~Pollmann}, \bibinfo{title}{Stark many-body localization}, \bibinfo{journal}{\href{http://dx.doi.org/10.1103/PhysRevLett.122.040606}{Phys. Rev. Lett.}} \href{http://dx.doi.org/10.1103/PhysRevLett.122.040606}{{\bf \bibinfo{volume}{122}}, \bibinfo{pages}{040606}}  (\href{http://dx.doi.org/10.1103/PhysRevLett.122.040606}{\bibinfo{year}{2019}}).

\end{thebibliography}

\end{document}